\documentclass[pre,twocolumn,preprintnumbers,amsmath,amssymb,showpacs,nofootinbib,floatfix]{revtex4}

\usepackage{graphicx,bm}

\makeatletter
\def\graphicscale{\twocolumn@sw{0.3}{0.4}}
\def\graphicthreescale{\twocolumn@sw{0.3}{0.4}}

\begin{document}

\title{Off-equilibrium scaling behaviors driven by 
time-dependent external fields \\

in three-dimensional O$(N)$ vector models}

\author{Andrea Pelissetto$^1$ and Ettore Vicari$^2$} 

\address{$^1$ Dipartimento di Fisica dell'Universit\`a di Roma ``La Sapienza"
        and INFN, Sezione di Roma I, I-00185 Roma, Italy}
\address{$^2$ Dipartimento di Fisica dell'Universit\`a di Pisa
        and INFN, Largo Pontecorvo 3, I-56127 Pisa, Italy}

\date{\today}

\begin{abstract}

We consider the dynamical off-equilibrium behavior of the
three-dimensional O$(N)$ vector model in the presence of a
slowly-varying time-dependent spatially-uniform magnetic field ${\bm
  H}(t) = h(t)\,{\bm e}$, where ${\bm e}$ is a $N$-dimensional
constant unit vector, $h(t)=t/t_s$, and $t_s$ is a time scale, at
fixed temperature $T\le T_c$, where $T_c$ corresponds to the
continuous order-disorder transition. The dynamic evolutions start
from equilibrium configurations at $h_i < 0$, correspondingly $t_i <
0$, and end at time $t_f > 0$ with $h(t_f) > 0$, or vice versa. We
show that the magnetization displays an off-equilibrium scaling
behavior close to the transition line ${\bm H}(t)=0$. It arises from
the interplay among the time $t$, the time scale $t_s$, and the finite
size $L$. The scaling behavior can be parametrized in terms of the
scaling variables $t_s^\kappa/L$ and $t/t_s^{\kappa_t}$, where
$\kappa>0$ and $\kappa_t > 0$ are appropriate universal exponents,
which differ at the critical point and for $T < T_c$. In the latter
case, $\kappa$ and $\kappa_t$ also depend on the shape of the lattice
and on the boundary conditions. We present numerical results for the
Heisenberg ($N=3$) model under a purely relaxational dynamics.  They
confirm the predicted off-equilibrium scaling behaviors at and below
$T_c$.  We also discuss hysteresis phenomena in round-trip protocols
for the time dependence of the external field.  We define a scaling
function for the hysteresis loop area of the magnetization that can be
used to quantify how far the system is from equilibrium.
\end{abstract}

\pacs{64.70.qj,64.60.Ht,64.60.an}

\maketitle



\section{Introduction}
\label{intro}

Statistical systems show notable off-equilibrium behaviors at phase
transitions.  For example, metastability and hysteresis phenomena
occur at first-order transitions~\cite{Binder-87}, while the
Kibble-Zurek (KZ) mechanism~\cite{Kibble-76,Zurek-85} is observed at
continuous transitions. They may arise when one of the model
parameters, such as the temperature or the external magnetic field in
spin systems, varies across the transition point with a time scale
$t_s$.  In this case some large-scale modes do not equilibrate, even
in the limit of large $t_s$, giving rise to peculiar off-equilibrium
behaviors.  These phenomena are of great interest in many different
physical contexts, see, e.g.,
Refs.~\cite{Binder-87,Kibble-76,Zurek-85,CG-04,PSSV-11,Biroli-15,DWGGP-16}.

Off-equilibrium phenomena due to the inability of the system to adapt
itself to changes of the external parameters have been much
investigated at continuous transitions.  As it happens for the
equilibrium static and dynamic critical behavior, off-equilibrium
behaviors driven by slow changes of external parameters show universal
features. For $t_s\to \infty$ and close to the transition point, one
can define general scaling expressions in terms of the standard
critical exponents characterizing the statics and dynamics of the
system at
equilibrium~\cite{Kibble-76,Zurek-85,GZHF-10,CEGS-12,Biroli-15}.  A
prototypical example is the KZ mechanism~\cite{Kibble-76,Zurek-85} for
the formation of topological defects when the temperature is slowly
changed across a continuous transition, from the disordered to the
ordered phase.  In this case, if $T_c$ is the critical temperature,
one considers the protocol $T(t)/T_c = 1 - t/t_s$ starting from
$t=t_i<0$ to $t=t_f>0$, where $t_s$ controls the speed of the
temperature variation.  In the large-$t_s$ limit, the system shows
off-equilibrium scaling behaviors across the
transition~\cite{Zurek-85,CEGS-12}.  Analogous phenomena are expected
at quantum transitions, when the system is driven across a continuous
transition by quasi-adiabatic changes of external parameters
\cite{KZ-q,PG-08,PSSV-11,CNGS-13,FDGZ-15}.  Many experiments have
addressed the same issues in several different physical systems
~\cite{CDTY-91,BCSS-94,BBFGP-96,Ruutu-etal-96,CPK-00,CGMB-01,MMR-02,
  MPK-03,CGM-06,MMARK-06,SHLVS-06,WNSBDA-08,GPK-10,CWCD-11,
  Chae-etal-12,MBMG-13,EH-13,Ulm-etal-13,Pyka-etal-13,LDSDF-13,
  Corman-etal-14,NGSH-15,Braun-etal-15}. We mention, as an example,
the recent studies characterizing the dynamic formation of
Bose-Einstein condensates, see, e.g., Ref.~\cite{DWGGP-16} and
references therein.

In this paper we consider off-equilibrium phenomena driven by
slowly-varying external fields coupled to the order parameter at
finite-temperature phase transitions, in systems characterized by a
continuous O($N$) symmetry.  In particular, we study the
off-equilibrium behavior of the three-dimensional (3D) O$(N)$ vector
model at fixed temperature $T$ in the presence of a slowly-varying
time-dependent spatially-uniform external field coupled to the vector
order parameter.  We consider a {\em magnetic} field ${\bm H}(t) =
h(t){\bm e}$ with fixed direction ${\bm e}$ and time-dependent
amplitude $h(t)=t/t_s$, where $t_s$ is a time scale.  In the
high-temperature paramagnetic phase, in which the correlation length
is finite in the infinite-volume limit, the system always reaches
equilibrium for sufficiently large time scales $t_s$.  Instead, at the
critical point $T_c$, infinite-volume systems are unable to
equilibrate for $t\approx 0$, when the magnetic field crosses the
transition point $h=0$, even in the large-$t_s$ limit.  The
off-equilibrium behavior close to the transition point turns out to be
universal. Its general features can be derived by using scaling
arguments analogous to those leading to the KZ
mechanism~\cite{GZHF-10,CEGS-12}.

Our main results concern the extension of these off-equilibrium
studies to the low-temperature $T<T_c$ phase, where a
spatially-uniform magnetic field drives first-order transitions: the
magnetization has a discontinuity at the transition point ${\bm H} =
0$ in the thermodynamic limit. As expected, equilibration becomes
significantly harder across first-order transitions.  However, we show
that slow variations of the magnetic field give rise to universal
off-equilibrium scaling behaviors also across this class of
first-order transitions, as in the continuous case.
General scaling predictions can be derived by combining the KZ
arguments with the general results for the equilibrium static and
dynamic behaviors of finite-size systems at first-order transitions.

Therefore, for any $T \le T_c$, and for $t\approx 0$ corresponding to
$h(t)\approx 0$, the dynamics of the system is expected to show a
universal off-equilibrium scaling behavior, which arises from the
interplay among the time $t$, the time scale $t_s$, and the size $L$
of the system. In this regime the time dependence of the magnetization
can be expressed in terms of scaling functions that depend on the
scaling variables $t_s^\kappa/L$ and $t/t_s^{\kappa_t}$, where
$\kappa>0$ and $\kappa_t > 0$ are appropriate universal exponents,
which differ at the critical point and for $T < T_c$. In the latter
case, $\kappa$ and $\kappa_t$ also depend on the shape of the lattice
and on the boundary conditions.  In order to check the general scaling
theory, we present a numerical analysis of the 3D Heisenberg ($N=3$)
model both for $T = T_c$ and for a few values of $T < T_c$.  In our
Monte Carlo (MC) simulations we consider a purely relaxational
dynamics.

We also extend our study to round-trip protocols, in which the
magnetic-field amplitude $h(t)$ is varied between $h_i < 0$ and
$h_f>0$ and then back again to $h_i < 0$. In this case we observe
off-equilibrium hysteresis phenomena at the critical point $T_c$ and
below $T_c$, which are characterized in terms of appropriate scaling
functions.

The paper is organized as follows.  In Sec.~\ref{offequi} we put
forward the general scaling theory appropriate to describe the
off-equilibrium phenomena occurring in the O($N$) vector model in the
presence of a slowly-varying magnetic field, both for $T=T_c$ and
$T<T_c$.  Sec.~\ref{eqbeh} discusses some features of the equilibrium
static and dynamic finite-size scaling of the O($N$) vector model in
the low-temperature phase $T<T_c$ and at $T_c$. The corresponding
critical exponents allow us to determine the appropriate scaling
variables that parametrize the off-equilibrium behavior of the
magnetization. In Sec.~\ref{numerical} we check the scaling arguments
by numerical simulations of the 3D Heisenberg lattice model under a
relaxational dynamics. We show that the magnetization displays the
predicted scaling behavior along the low-temperature first-order
transition line ($T<T_c$) and at the critical point $T_c$.  In
Sec.~\ref{hysteresissec} we discuss the hysteresis phenomena that
occur in round-trip protocols, in which the external field is first
increased from $h_i< 0$ to $h_f>0$ and then decreased again to
$h_i$. We determine the scaling behavior of the area enclosed by the
hysteresis loop of the magnetization at and below $T_c$.  Finally, in
Sec.~\ref{conclu} we draw some conclusions.

\section{Off-equilibrium scaling driven by magnetic fields}
\label{offequi}

\subsection{The 3D O($N$) vector model}
\label{onsec}

We study the 3D O($N$) vector model in the presence of a uniform
magnetic field.  We consider $N$-component unit-length spins ${\bm
  s}_i$, defined on a simple cubic lattice, and the Hamiltonian
\begin{equation}
{\cal H} = - J \sum_{\langle ij \rangle} {\bm s}_i \cdot {\bm s_j} - {\bm
  H}\cdot\sum_i {\bm s}_i,
\label{ham}
\end{equation}
where $\langle ij \rangle$ indicates nearest-neighbor sites and $J>0$.  In the
following we set $J=1$, so that all energies are expressed in units of $J$.
We write the magnetic field as
\begin{equation}
{\bm H} = h{\bm e},   \qquad {\bm e}\cdot {\bm e}=1.
\label{hh}
\end{equation}
We consider cubic ($L\times L\times L$) and anisotropic cylinder-like
($L\times L \times L_\parallel$ with $L_\parallel\propto L^2$)
systems, with periodic boundary conditions (PBC) along all directions.

In the absence of an external magnetic field, i.e., for ${\bm H}=0$,
the system undergoes a continuous transition at a finite temperature
$T_c$, which separates the high-temperature paramagnetic phase from
the low-temperature ferromagnetic phase.  In the low-temperature phase
$T<T_c$, the external magnetic field drives first-order transitions at
$h=0$, giving rise to a discontinuity in the magnetization. We have
\begin{equation}
\lim_{h\to \pm 0} \, \lim_{L\to\infty} m(L,h,T) = \pm\, m_0(T),
\label{m0t}
\end{equation}
where
\begin{equation}
m \equiv {\bm M}\cdot {\bm e}, \qquad {\bm M} = {1\over V} \langle
\sum_i {\bm s}_i \rangle.
\label{mdef}
\end{equation}
The spontaneous magnetization $m_0(T)$ varies from $m_0=1$ for $T= 0$
to $m_0\to 0$ for $T\to T_c$. Approaching the critical point, the
magnetization $m_0(T)$ behaves as $m_0(T)\sim (T_c-T)^\beta$, where
$\beta$ is the magnetization critical exponent.

\subsection{Off-equilibrium protocol}
\label{protocol}

We are interested in the off-equilibrium dynamics arising in the
presence of a time-dependent spatially-uniform magnetic field ${\bm
  H}(t)$ crossing the transition point ${\bm H}=0$ at fixed
temperature $T$.  We assume that the direction of the magnetic field
(vector ${\bm e}$) is fixed in the dynamics, while the component along
$\bm e$ varies as
\begin{equation}
h(t) = t/t_s,
\label{ht}
\end{equation}
where $t_s$ is a time scale and we have chosen $t$ so that $t=0$
corresponds to $h=0$.  In the off-equilibrium protocol one starts from
equilibrium configurations at an initial value $h_i<0$ at time
$t_i<0$. Then, the magnetic field is slowly changed up to a time
$t_f>0$, corresponding to a finite $h_f>0$.  This procedure is
repeated several times, starting the dynamics from different
equilibrium configurations at $h = h_i$. Observables are then averaged at fixed
time $t$.  As we shall see, in the limit $t_s\to \infty$ (very slow
dynamics) the off-equilibrium scaling behavior around $t=0$ is
universal and does not depend on the initial value $h_i<0$ of the
magnetic field.

One can also consider a reversed process in which the magnetic field
(\ref{ht}) varies from $h_i>0$ to $h_f<0$. Formally, it can be
obtained by decreasing the {\em time} parameter from $t_i>0$ to
$t_f<0$.  This new process is of course identical to the original one,
provided one changes the sign of $t$, $h$, and of the magnetization.

The off-equilibrium behavior depends on the particular time evolution
of the system.  There are several physically interesting cases, see,
e.g., Refs.~\cite{HH-77,FM-06}. In the following we present general
scaling arguments that apply to any dynamics in which the slowest mode
is associated with the magnetization.  They will be verified
numerically in Sec.~\ref{numerical} for a purely relaxational
dynamics.

\subsection{Off-equilibrium scaling across the transitions}
\label{offequisec}

We present a scaling theory for the off-equilibrium dynamics of the
magnetization across $h=0$, in the quasi-adiabatic limit, i.e. for
$t_s\to \infty$.  In off-equilibrium processes driven by spatially
uniform external fields, low-momentum modes, and in particular the
magnetization, are expected to be the slowest modes of the system.
The size dependence of the relevant scaling variables is parametrized
by the effective RG scaling dimension $y_h$ of the magnetic field and
by the dynamic exponent $z_m$ associated with the equilibrium dynamics
of the magnetization.  The determination of the appropriate values of
$y_h$ and $z_m$ for the cases we consider, i.e., cubic and anisotropic
cylinder-like systems, in the low-temperature phase and at $T_c$, is
postponed to Sec.~\ref{eqbeh}.

The scaling theory is meant to describe the deviations of the
statistical correlations from their equilibrium value, due to the fact
that the system is not able to adapt itself to the changes of the
magnetic field across ${\bm H}=0$.  Assuming the existence of a
nontrivial scaling behavior for $h(t)\approx 0$ ($t\approx 0$,
correspondingly), we expect the off-equilibrium behavior to be
controlled by the two scaling variables
\begin{eqnarray}
r_1 = h(t)L^{y_h}=(t/t_s)L^{y_h},\qquad  r_2 = t L^{-z_m}.
\label{s12}
\end{eqnarray}
In our context, it is convenient to use the equivalent scaling variables
\begin{eqnarray}
&u \equiv  {t_s^\kappa/L},\qquad
&\kappa = {1\over y_h + z_m} ,
\label{udef}\\
&w \equiv  {t / t_s^{\kappa_t}},\qquad
&\kappa_t = z_m \kappa,
\label{wdef}
\end{eqnarray}
which are combinations of $r_1$ and $r_2$: $u = (r_2/r_1)^\kappa$ and
$w = (r_1/r_2)^{\kappa_t} r_2$. 
Note that  generally $\kappa_t < 1$.

We consider here a magnetic field that varies linearly with
$t$, but one could analogously consider nonlinear behaviors such as
$h(t) = {\rm sgn}(t)\;|t/t_s|^{n}$ for $n>0$.  In this case the
relevant scaling variables should be $r_1=h(t) L^{y_h}$ and
$r_2=tL^{-z_m}$, which could be replaced by $u = t_s^\kappa/L$ and $w
= t/t_s^{\kappa_t}$, with exponents $\kappa = n/(y_h + n z_m)$ and
$\kappa_t = z_m \kappa$.

We now assume that the dynamics across the transition presents a
scaling behavior when $L,\,t_s$, and $t$ become large at fixed $u$ and
$w$.  This implies the emergence of a length scale $\xi\sim
t_s^\kappa$ and of a time scale $\tau\sim t_s^{\kappa_t}$ across the
transition. The equilibrium static finite-size scaling (FSS) should be
recovered in the limit $u,|w| \to\infty$ keeping $r_1=w/u^{y_h}$
fixed.  The off-equilibrium scaling behavior does not depend on the
choice of the initial $h_i$ and of the final $h_f$, because scaling
occurs in a narrow range of values of $|h|$ that shrinks as $t_s\to
\infty$.  Indeed, the scaling behavior is observed in a time interval
of size $\tau\sim t_s^{\kappa_t}$ around $t=0$.  Since $\kappa_t<1$,
we have $\tau/t_s\to 0$ in the large-$t_s$ limit.  Therefore, scaling
occurs for smaller and smaller values of $|h|$. Note that this argument
applies for any $T\le T_c$.

Similar off-equilibrium scaling arguments have been reported in
Refs.~\cite{GZHF-10,CEGS-12,CPV-15,PV-15} to describe other
off-equilibrium systems driven by slowly-varying model parameters at
phase transitions, such as the Kibble-Zurek mechanism.  Actually,
these off-equilibrium behaviors may be exploited to determine the
equilibrium static and dynamic properties of critical systems, see,
e.g., Ref.~\cite{LPSY-15}.

Let us now derive more quantitative predictions, considering first the
dynamics at the critical point.  The equilibrium magnetization
(\ref{mdef}) satisfies the scaling relation
\begin{eqnarray}
m(L,h,T_c) \approx L^{-y_m} f_m(hL^{y_h}),\label{mefss}
\end{eqnarray}
where 
\begin{equation}
y_m = 3-y_h = {1+\eta\over 2},\qquad y_h = {5-\eta\over 2},
\label{ymyhct}
\end{equation}
$y_m$ is the RG dimension of the order parameter, and $\eta$ is the
critical exponent parametrizing the short-distance behavior of the
two-point function.  The behavior of $f_m(x)$ for large values of
$|x|$ can be obtained by requiring Eq.~(\ref{mefss}) to be consistent 
with the finite-$h$ infinite-volume behavior of the magnetization
\cite{PV-02},
\begin{equation}
m(L=\infty,h,T_c) \approx \pm a |h|^{y_m/y_h}\quad {\rm for}\;\;
|h|\to 0, 
\label{mltce}
\end{equation}
where $a > 0$ and the $\pm$ sign reflects the fact that $m$ changes
sign as $h\to -h$.  Requiring the scaling formula (\ref{mefss}) to
reproduce Eq.~(\ref{mltce}) in the limit $L\to \infty$ for 
small nonvanishing values of $|h|$, we obtain for
$|x| \to \infty$
\begin{equation}
f_m(x) \approx f_{\pm\infty}(x) \equiv  \pm a |x|^{y_m/y_h}.
\label{fm-largeL}
\end{equation}
Indeed, if we substitute $f_m(x)$ with $f_{\pm\infty}(x)$ in
Eq.~(\ref{mefss}) and use $x= hL^{y_h}$, we reobtain
Eq.~(\ref{mltce}).

In order to describe the off-equilibrium regime around the transition
point, we generalize Eq.~(\ref{mefss}) by writing
\begin{equation}
m(t,t_s,L;T=T_c) \approx L^{-y_m} F_m(u,w).
\label{mcont}
\end{equation}
The {\em thermodynamic} infinite-volume limit (before taking the large
$t$ and $t_s$ limits) can be formally obtained by performing the limit
$u\to 0$ keeping $w$ fixed, which leads to scaling expressions
analogous to those discussed in Ref.~\cite{CEGS-12}.  Note that
$F_m(u,w)$ is not expected to be symmetric for $w\to -w$ for finite
values of $u$, because the off-equilibrium process is irreversible and
therefore the time-reversal symmetry is violated (it is only recovered
in the static FSS limit and for $|w|\to\infty$, see below).

The limit $|w|\to\infty$ of $F_m(u,w)$ at fixed $u$ is expected to
lead to the infinite-volume equilibrium behavior. Indeed, note first
that, in a finite volume $L$, the slowest time scale --- in our system
the autocorrelation time $\tau_m$ associated with the magnetization
--- scales as $L^{z_m}$. A necessary condition to obtain equilibrium
results is therefore that $t_s \gg \tau_m$, i.e., $t_s L^{-z_m} \to
\infty$. At fixed $u$ we have $t_s L^{-z_m} = u^{1/\kappa} L^{y_h}$
and hence the condition is satisfied for $L\to \infty$. Since we take
the limit $w\to\infty$, we are considering the system at times $t$
much larger than the time scale at which the off-equilibrium behavior
occurs, so that the system is in equilibrium. Therefore, the scaling
function $F_m(u,v)$ should match its equilibrium counterpart
$f_m(r_1)$. Finally, since $r_1 = h L^{y_h} = w u^{-y_h}$, in the
limit $w\to\infty$ at fixed $u$ we have $r_1 \to \infty$, i.e., we are
considering the behavior in the infinite-volume limit. Therefore, we
expect
\begin{eqnarray}
F_m(u,w) \approx f_{+\infty}(w u^{-y_h}) &= &
   a (w u^{-y_h})^{y_m/y_h} \nonumber \\
  &=& a\,u^{-y_m}  \,w^{y_m/y_h} .
\label{F-largew}
\end{eqnarray}
In our discussion, we also need the corrections to the behavior
(\ref{F-largew}).  We now argue that the approach is exponentially
fast. Note that, for $w\to \infty$, we are investigating the behavior
of the system at finite (small) values of the magnetic field.  For
finite $h$ and in infinite volume, we expect the deviations from
equilibrium to decay exponentially in $t$ with a typical time scale
$\tau_h$, i.e., as $\sim t^b e^{-t/\tau_h}$, where we have also included a
power correction with exponent $b$.  The time scale $\tau_h$ should be
of order $\xi_h^{z_m}$ and the correlation length $\xi_h$ should scale
as $h^{-1/y_h}$. In the scaling limit at fixed $w$ we have 
\begin{equation}
{t\over \tau_h} \sim t h^{z_m/y_h} = t^{z_m/y_h + 1} t_s^{-z_m/y_h} = 
    t^\varepsilon t_s^{-\kappa_t \varepsilon} = w^\varepsilon,
\end{equation}
where $\varepsilon = 1 + z_m/y_h=1/(y_h\kappa)>0$. Therefore, for large values 
of $w$, we expect
\begin{equation}
{F_m(u,w)\over f_{+\infty}(w u^{-y_h})}  - 1 \sim w^b \exp(-c \,w^\varepsilon),
\label{asyexp}
\end{equation}
where $c$ should depend on $u$.

The same argument can be used to discuss the limit $w\to-\infty$.  In
this case, we start the dynamics at a very large negative time $t_i$
with $|t_i| \sim t_s$ in the magnetized phase with $h < 0$. Hence the
previous arguments should apply also here, so that we predict, for $w
\to -\infty$ at fixed $u$, the behavior
\begin{equation}
{F_m(u,w)\over f_{-\infty}(w u^{-y_h})}  - 1 \sim |w|^b \exp(-c\,|w|^\varepsilon),
\label{asyexp2}
\end{equation}
Notice that the constant $c$ and the exponent $b$ entering here might
well differ from those appearing in Eq.~(\ref{asyexp}), while the
exponent $\varepsilon$ is expected to be the same.

The off-equilibrium scaling behavior (\ref{mcont}) can be
straightforwardly extended to other observables and correlation
functions.  Moreover, small deviations from $T_c$ can be taken into
account by adding a further dependence on the scaling variable
$(T-T_c)L^{1/\nu}$.

The emergence of a universal static FSS behavior has been established
also in the case of first-order transitions, see, e.g.,
Refs.~\cite{PF-83,FP-85,CNPV-14}. In particular, Eq.~(\ref{mefss})
also holds at discontinuous transitions (a detailed discussion in
presented in Sec.~\ref{fssmtc}). It is therefore natural to conjecture
that the off-equilibrium ansatz (\ref{mcont}), with the appropriate
exponents, applies also along the low-temperature first-order
transition line.  Since $m \to \pm m_0(T)$ for $h\to 0^\pm$,
cf. Eq.~(\ref{m0t}), Eq.~(\ref{fm-largeL}) formally requires $y_m = 0$
and $a=m_0(T)$.  To avoid an additional nonuniversal normalization
constant, it is convenient to introduce a {\em renormalized}
magnetization defined as
\begin{eqnarray}
m_r(t,t_s,L;T) \equiv {m(t,t_s,L;T)\over m_0(T)}, 
\label{renmag}
\end{eqnarray}
which satisfies $-1\le m_r \le 1$.  Therefore, in the low-temperature 
phase and for $t_s\to\infty$, 
we expect --- we will verify it numerically below --- the 
scaling behavior
\begin{eqnarray}
m_r(t,t_s,L;T) \approx F_m(u,w) , \label{fm}
\end{eqnarray}
where $F_m$ is a universal scaling function, which takes values in the
range $-1\le F_m \le 1$ and satisfies all scaling expressions reported
above. Using arguments analogous to those leading to
Eqs.~(\ref{asyexp}) and (\ref{asyexp2}), we expect an exponential
approach to the equilibrium behavior in the $w\to\pm \infty$ limits,
i.e.
\begin{eqnarray}
1 - F_m(u,w)   &\sim& e^{-c_+ \,|w|^\varepsilon} \quad  {\rm for}\;\;
w\to \infty, \label{asyexp3p}\\
-1 - F_m(u,w)   &\sim& - e^{-c_- \,|w|^\varepsilon} \quad {\rm for}\;\;
w\to -\infty, \nonumber 
\end{eqnarray}
where again $\varepsilon = 1 +z_m/y_h$, and
we have neglected possible powers of $w$ in the prefactor.

We expect universality with respect to temperature changes in the
low-temperature phase. This is supported by the fact that the
effective scaling dimensions $y_h$ and $z_m$ do not depend on
temperature for $T<T_c$ (but they differ from those at $T_c$), see
sections \ref{fssmtc} and \ref{dynmtc}.  Universality also implies
that the scaling function $F_m$ does not depend on $T$, apart from a
trivial renormalization of the scaling variables $u$ and $w$.

One can also consider the time-reversed protocol, in which one starts
at $h_i > 0$ and then decreases $h(t)$ to $h_f<0$, see
Sec.~\ref{protocol}.  If $F_m^{({\rm inv})}(u,w)$ is the corresponding 
scaling function appearing in Eqs.~(\ref{mcont}) and (\ref{fm}), 
the symmetry of the model with respect to a
reflection of the magnetic field implies 
\begin{equation}
F_m^{({\rm inv})}(u,w) = - F_m(u,-w). 
\label{fminv}
\end{equation}

\section{Equilibrium static and dynamic behavior}
\label{eqbeh}

To verify the off-equilibrium scaling behaviors put forward in the
previous section, we consider the Heisenberg $N=3$ model. It presents
a continuous transition at~\cite{BFMM-96,BC-97} $T_c \approx 1.443$
(Ref.~\cite{BFMM-96} obtained $\beta_c\equiv 1/T_c =
0.69300(1)$ by a FSS analysis of MC data), which separates the
low-temperature ferromagnetic phase, in which the symmetry is
spontaneously broken, from the high-temperature paramagnetic phase. In
this section we review some general results for the static scaling
behavior in a finite volume, both at $T=T_c$ and for $T < T_c$. We
also discuss the dynamic equilibrium behavior for a relaxational
dynamics, focusing, in particular, on the low-temperature phase.

\subsection{Equilibrium finite-size scaling}
\label{fssbeh}

\subsubsection{Finite-size scaling at the critical point}
\label{fsstc}

At the critical point, the model shows a universal FSS depending on
the static critical exponents.  For the Heisenberg universality class
they have been accurately estimated by various methods: see, e.g.,
Ref.~\cite{PV-02} for a review of results. We mention the
estimates~\cite{HV-11,CHPRV-02} $\nu=0.7117(5)$ and $\eta= 0.0378(3)$
for the correlation-length and two-point function exponents, from
which one can derive the magnetization exponent $\beta=\nu
(1+\eta)/2$.

At $T_c$, the magnetization behaves as reported in Eq.~(\ref{mefss}),
with $y_m=0.5189(2)$ and $y_h=2.4811(2)$.  The scaling function $f_m$
is universal apart from trivial normalizations. However, it depends on
the shape of the system and on the boundary conditions (but the
critical exponents remain unchanged). Scaling corrections decay as
$L^{-\omega}$, where the exponent $\omega$ is related to the RG
dimension of the leading irrelevant operator \cite{PV-02}.  Estimates
of $\omega$ by various methods give $\omega\approx
0.78$~\cite{GZ-98,Hasenbusch-01}.

\subsubsection{Finite-size scaling at the first-order transition line}
\label{fssmtc}

\begin{figure}[tbp]
\includegraphics*[scale=\graphicscale]{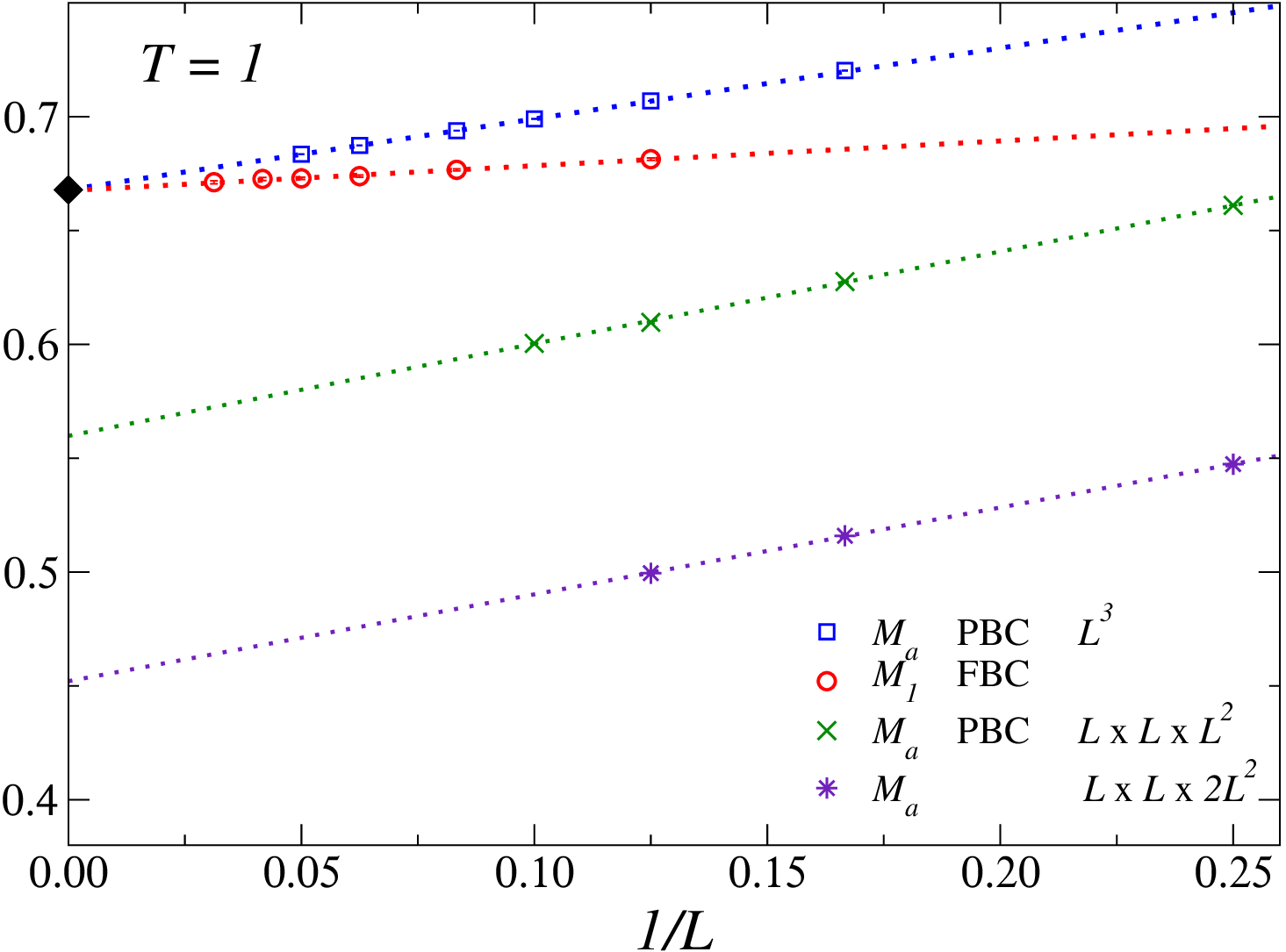}
\caption{(Color online) Finite-size magnetization in cubic $L^3$
  systems and in anisotropic $L\times L \times L_\parallel$ systems
  for $T=1$ and ${\bm H}=0$.  In the cubic case ($L_\parallel = L$) we
  report $M_a$ defined in Eq.~(\ref{calm}) for systems with PBC, and
  the component $M_1 = {\bm M}\cdot {\bm e}$ of the magnetization for
  systems with fixed boundary conditions (FBC): ${\bm s}_i = {\bm e} =
  (1,0,0)$ on the boundary.  The finite-size results approach the same
  value $m_0(T)$ as $L\to\infty$, with $L^{-1}$ corrections.  The
  dotted lines correspond to linear fits to $M_\infty+b/L$, which lead to the
  large-$L$ extrapolated estimates $M_{a,\infty} = 0.66804(4)$ and
  $M_{1,\infty}=0.6676(6)$.  In the anisotropic case we report $M_a$
  for $L_\parallel = L^2$ and $L_\parallel = 2 L^2$.  They also
  approach finite values $M_{a,\infty}<m_0(T)$, which depend on the
  ratio $L^2/L_\parallel$. Corrections are of order $L^{-1}$.  }
\label{magh0}
\end{figure}

FSS in the low-temperature phase is more complex, because the relevant
scaling variables depend both on the shape of the system and on the
boundary conditions \cite{FP-85,CNPV-14,BDT-00}.  Fig.~\ref{magh0}
shows some numerical results obtained by standard MC simulations at
$T=1$ and ${\bm H}=0$, for cubic $L^3$ systems with PBC. Since ${\bm
  H}=0$, we have ${\bm M}=0$ by symmetry. However, the related
quantity
\begin{equation}
M_a = {1\over V} \langle | \sum {\bm s}_i | \rangle 
\label{calm}
\end{equation}
does not vanish, even in the large-volume limit.  We compare the
estimates of $M_a$ with those of $M_1 = {\bm M\cdot \bm e}$ [here we
  take ${\bm e} = (1,0,0)$] computed in cubic systems with fixed
boundary conditions (FBC) (${\bm s}_i = \bm e$ on the boundary), whose
large-$L$ limit is expected to provide $m_0(T)$, cf. Eq.~(\ref{m0t}).
As shown in Fig.~\ref{magh0}, the extrapolated value of $M_a$ is
consistent with $m_0(T)$.  For example, linear fits of $M_a$ and $M_1$
at $T=1$ to $M_{\infty} +b/L$ give $M_{a,\infty}= 0.66804(4)$ (PBC
case) and $m_0=M_{1,\infty} = 0.6676(6)$ (FBC case).  These results
confirm the arguments of Ref.~\cite{FP-85}. They argued that in cubic
systems the modulus of the magnetization is essentially constant and
asymptotically equal to $m_0(T)$. On the other hand, its direction is
randomly distributed in the sphere, to recover the O($N$) invariance.
Spin waves give only rise to subleading contributions that decrease as
$1/L$~\cite{FP-85}.  We also note that $m_0(T)$ increases with
decreasing $T$ (we expect $m_0(T)\to 1$ for $T\to 0$).  For example,
at $T=0.5$ we obtain $m_0(T) =0.8600(3)$.

In the case of cubic systems with PBC the effective RG dimension $y_h$ 
of the magnetic field is $y_h=d=3$~\cite{FP-85}.
Thus, we expect \cite{FP-85}
\begin{eqnarray}
&&m(h,L;T) = m_0(T) f_m(hL^3),\quad f_m(0) = 0 , 
\label{nthlfo}\\
&&M_a(h,L;T) = m_0(T) f_a(hL^3),\quad f_a(0) = 1.
\label{fssmthl}
\end{eqnarray}
In particular, Ref.~\cite{FP-85} obtains $f_m(h L^3) = I_{3/2}(y)/I_{1/2}(y)$
for $N=3$, 
where $I_\nu(y)$ are the modified Bessel functions of the first kind
and $y=m_0 h L^3$.  Corrections, such as those due to spin-wave modes,
are expected to decay as $1/L$~\cite{FP-85}.

FSS is expected to substantially change in anisotropic geometries. If
we consider systems of size $L\times L \times L_\parallel$ and
$L_\parallel \gg L$, a finite longitudinal length scale~\cite{FP-85}
emerges along the longitudinal direction, which scales as
$\xi_\parallel \sim L^2$.  More precisely, for cylinder-like systems
one can write the asymptotic scaling relation at $h=0$ as
\begin{equation}
\xi_\parallel \approx L^2 f_\xi(L_\parallel/L^2),
\label{xiparsca}
\end{equation}
where $f_\xi(x)$ is a scaling function, independent of $T$ apart from
trivial multiplicative normalizations.  This implies that the
effective scaling dimension of $h$ in the FSS of $L\times L \times
L_\parallel$ systems with $L_\parallel\sim L^2$ is $y_h=d+1=4$. The
FSS of the observables related to the magnetization reads~\cite{FP-85}
\begin{eqnarray}
m(h,L;T) &\approx& m_0(T) \,g_m(hL^4,L_\parallel/L^2),
\label{mscasta}\\
M_a(h,L;T) &\approx& m_0(T)\,  g_a(hL^4,L_\parallel/L^2),
\label{mascasta}
\end{eqnarray}
where $g_{m,a}(x,y)$ are scaling functions. In particular, we expect
$g_a(0,y\to 0) = 1$ and $g_a(0,y\to\infty)=0$.  These scaling
behaviors are confirmed by the numerical results shown in
Fig.~\ref{magh0} for $M_a$ at $T=1$ and $h=0$, for 
$L_\parallel = L^2$ and $L_\parallel = 2
L^2$. Note that the asymptotic value of $M_a$
decreases with increasing $L_\parallel/L^2$.

\subsection{Equilibrium dynamic behavior under a relaxational dynamics}
\label{edfss}

The off-equilibrium behavior depends on the dynamic universality class
of the dynamics driving the system across the transition. It is 
generally characterized by the dynamic exponent $z$, which specifies
the equilibrium large-$L$ behavior of the autocorrelation time $\tau$
of the observables coupled to the slowest modes: $\tau\sim L^z$.

In the following we consider a purely relaxational dynamics, which can
be realized by using the standard heat-bath~\cite{heatbath} or
Metropolis~\cite{metropolis} algorithms in MC simulations.

\subsubsection{Equilibrium dynamics at the critical point}
\label{dyntc}

At the critical point $T_c$, the dynamic exponent for the relaxational
dynamics (model A of Ref.~\cite{HH-77}) is very close to two.  The
dynamic exponent has been computed to three loops in perturbation
theory~\cite{AV-84} (see also Refs.~\cite{HHM-72,HH-77,FM-06}); its
$\epsilon$ expansion can be expressed in terms of the exponent $\eta$
as
\begin{equation}
z=2 + c\,\eta,\qquad c = 0.726 - 0.137\,\epsilon+O(\epsilon^2).
\label{zcthree}
\end{equation}
Using the 3D estimate $\eta=0.0378(3)$, this relation leads to the
estimate $z=2.02(1)$ for the 3D Heisenberg universality class, where
the error takes somehow into account the uncertainty on the
extrapolation to $\epsilon=1$ of the three-loop $\epsilon$-expansion
result.  In model-A systems the dynamic exponent $z$ controls the
equilibrium critical behavior of all observables associated with the
critical modes, including the magnetization. Therefore, the exponent
$z_m$ entering the off-equilibrium scaling exponent $\kappa$ defined
in Eq.~(\ref{udef}) should be identified with $z$: $z_m=z=2.02(1)$.

Note that $z$ not only depends on the static universality class, but also 
on the type of dynamics \cite{HH-77}. For instance, if the total
magnetization is conserved in the dynamics, 
the critical dynamics for a Heisenberg ferromagnet is analogous to that 
of model J of Ref.~\cite{HH-77}, for which \cite{HH-77}
$z=y_h=2.4811(2)$.

\subsubsection{Equilibrium dynamics at the first-order transition line}
\label{dynmtc}

\begin{figure}[tbp]
\includegraphics*[scale=\graphicscale]{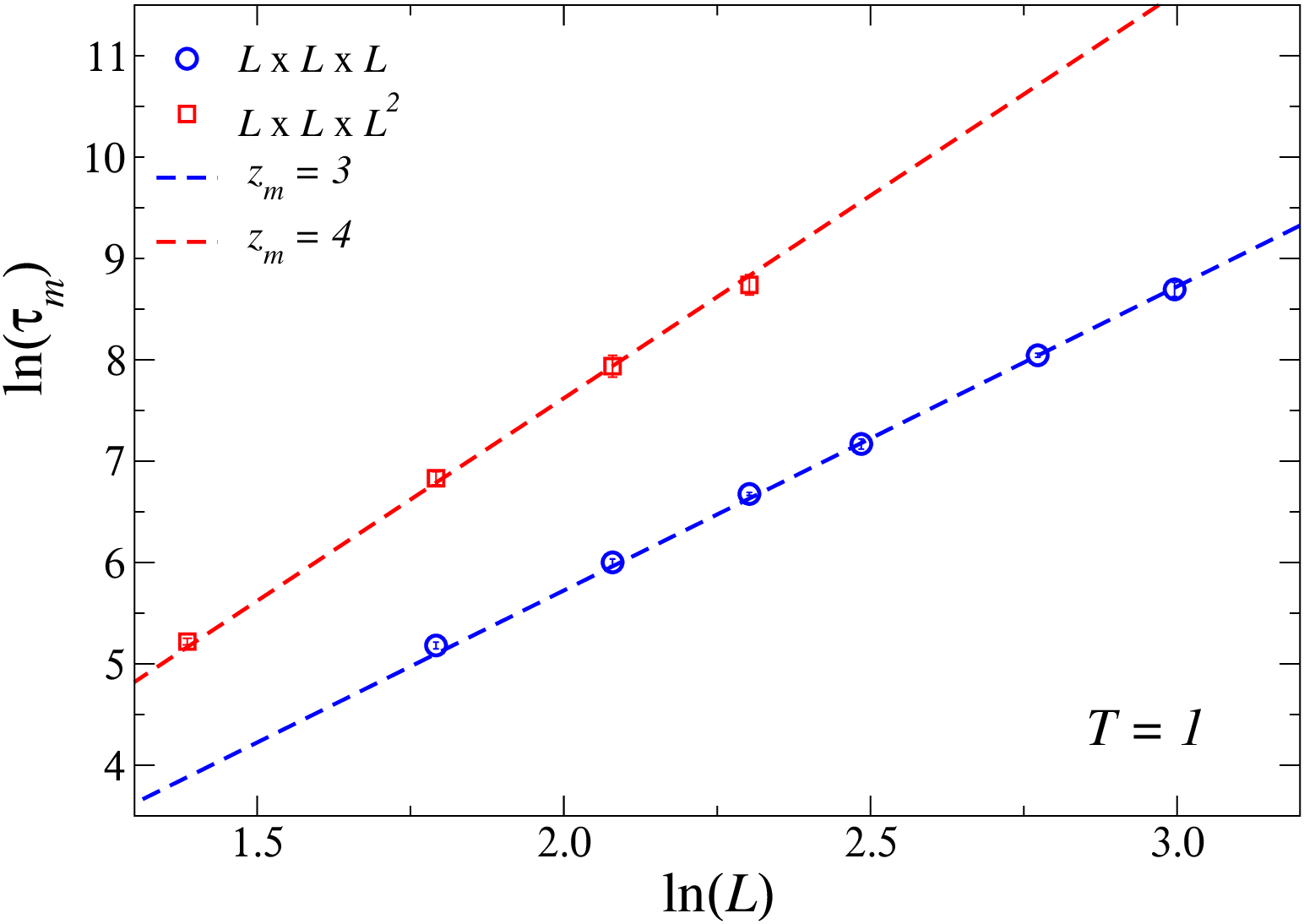}
\caption{(Color online) Autocorrelation time $\tau_m$ (in units of
  lattice sweeps) of the magnetization at $T=1$ and $h=0$ for the
  heat-bath dynamics~\cite{heatbath}. We consider cubic $L\times L
  \times L$ systems and anisotropic $L\times L \times L^2$ systems,
  with PBC in all directions.  The lines correspond to fits to
  $\tau_m=cL^{z_m}$, with $z_m= 3$ and $z_m= 4$ in the two cases,
  respectively.  }
\label{taum}
\end{figure}

Let us now consider the dynamic behavior in the low-temperature case.
For this purpose we perform MC simulations at $T=1$ using the
heat-bath dynamics~\cite{heatbath}.  Fig.~\ref{taum} shows
data~\cite{intauto} for the integrated autocorrelation time $\tau_m$
of the magnetization ${\bm M}$ defined in Eq.~(\ref{mdef}).  In the
case of cubic $L^3$ systems, a fit of all data ($L$ varies between 6
and 20) to $\tau_m=a L^{z_m}$ gives $z_m=2.92(4)$. If we only consider
data satisfying $L>10$, we obtain $z_m = 3.00(15)$. Results are
clearly consistent with $z_m = 3$: a fit of all data to $\tau_m=a L^3$
gives $\chi^2/{\rm d.o.f.}\approx 1.0$ where d.o.f. is the number of
degrees of freedom of the fit (the result is reported in
Fig.~\ref{taum}).

In the case of cylinder-like systems ($L\times L \times L^2$), 
a fit of the magnetization 
integrated autocorrelation time ($4\le L \le 10$) to $\tau_m=a L^{z_m}$ 
gives $z_m=3.9(1)$. This result is clearly larger than that obtained in the 
cubic case and is consistent with $z_m = 4$: a fit to $\tau_m=a L^4$
has $\chi^2/{\rm d.o.f.}\approx 0.8$ (see Fig.~\ref{taum}).
Also the
autocorrelation time of $M_a$ defined in Eq.~(\ref{calm}) apparently
increases as $L^4$, at variance  with the cubic case, in which the
autocorrelation time of $M_a$ is much smaller than that of $m$.
  
These numerical results suggest $z_m=3$ for
cubic $L^3$ systems and $z_m=4$ for anisotropic cylinder-like systems, when
PBC are considered. In the following we argue that they are exact
values. We shall first present a heuristic argument, then a calculation using 
the Langevin dynamics for the $\phi^4$ theory.

\subsubsection{Theoretical predictions for the dynamic behavior of the 
magnetization}

Let us first consider the case of fixed boundary conditions
breaking the O($N$) symmetry, obtained by 
fixing the direction of the spins on the boundaries or introducing a boundary
magnetic field. The average magnetization vector $\bm M$
is essentially fixed: its modulus is approximately equal to 
$m_0(T)$ and its direction is fixed by the boundary conditions. 
In this case, the only relevant modes 
are spin waves (Goldstone modes). Since there are no critical fluctuations,
the theory is expected to be Gaussian and therefore $z_m = 2$.

Let us now consider PBC or, more generally, any type of boundary
conditions that do not break the O($N$) invariance. This case is different
from that considered above. Indeed, while the modulus of the magnetization is 
again essentially fixed, there is no constraint on the direction 
of $\bm M$, whose rotation turns out to be the slowest dynamic mode of the 
system.  

As we discussed in Sec.~\ref{fssmtc}, in a finite volume the static
behavior depends on the shape of the system. In the cubic case, the
system is strongly correlated, so that all spins ${\bm s}_i$ point in
the same direction, with small random fluctuations. This implies that
the modulus of the magnetization ${\bm M}$ defined in Eq.~(\ref{mdef})
is essentially equal to $m_0(T)$ in the large-$L$ limit. The direction
of $\bm M$ is however not fixed. Because of the relaxational dynamics,
the vector ${\bm S}=\sum_i {\bm s}_i$ performs a random walk on the
sphere of radius $|{\bm S}|\sim m_0(T) L^d$.  Therefore, we expect
that the number of random movements of the local site variables ${\bm
  s}_i$, which are required to significantly move the vector ${\bm
  S}$, scales as $L^{2d}$.  This leads to the result $z_m=d$, taking
into account that the time unit, i.e., a complete sweep of the
lattice, consists of $L^d$ local updatings.

A similar argument shows that, in anisotropic $L\times L \times
L_\parallel$ systems with $L_\parallel/L^2$ fixed, the correct
exponent is $z_m=4$.  In this case we should take into account the
emergence of a longitudinal correlation length $\xi_\parallel\sim
L^2$, which measures the correlations among the directions of the
local magnetization (or its sum over $L^2$ slices) along the
longitudinal direction (the one with length $L_\parallel$).  As
discussed in Ref.~\cite{FP-85}, the local magnetization is not
uniformly directed along the longitudinal direction.  However, the
system can be partitioned in subsystems of size $L\times L \times
\xi_\parallel$, such that in each of them the spins ${\bm s}_i$ are
mostly oriented in the same direction, guaranteeing that the average
magnetization ${\bm \mu}$ of the subsystem satisfies $|{\bm \mu}|
\approx m_0(T)$.  Assuming that the slowest modes are those related to
the rotation of the average magnetization of each subsystem, the same
argument used for cubic systems implies that $\tau_m \sim L^2
\xi_\parallel \sim L^4$, so that $z_m=4$ for cylinder-like anisotropic
systems.

We may obtain these results in a more quantitative way, considering
the Langevin \cite{VanKampen}
dynamics in a O($N$) symmetric $\phi^4$ theory. We assume for
simplicity $N=2$, considering a complex one-component field $\phi$.
However, the results hold for any $N\ge 2$.  The Langevin equation
controlling the dynamics is given by \cite{HH-77,FM-06}
\begin{equation}
{\partial \phi({\bm x},t)\over \partial t} = 
-{\Omega\over2} 
\left(-\nabla^2 \phi + r \phi + {\lambda\over 6} \phi |\phi|^2 \right) +
\eta({\bm x},t),
\label{Langevin}
\end{equation}
where the random variables $\eta({\bm x},t)$ have a probability distribution
proportional to 
\begin{equation}
\exp\left[ - {1\over 2 \Omega} 
    \int dt d{\bm x} \, |\eta({\bm x},t)|^2\right].
\end{equation}
We will consider averages with respect to the random noise, which will be 
labelled with the angular brackets, i.e., with $\langle \cdot \rangle$.

In the low-temperature phase we have $r < 0$, the system is
magnetized, and $|\phi({\bm x},t)|$ fluctuates around
\begin{equation}
a = \sqrt{-6 r/\lambda}.
\end{equation}
Therefore, we write
\begin{equation}
\phi({\bm x},t) = a e^{i\theta({\bm x},t)} [1 + \delta({\bm x},t)],
\end{equation}
where $\delta({\bm x},t)$ is real and takes into account the
fluctuations of $|\phi|$. Inserting this expression into
Eq.~(\ref{Langevin}) we obtain
\begin{eqnarray}
&& i {\partial \theta\over \partial t} (1 + \delta) + 
  {\partial \delta\over \partial t} = 
\\
&& \qquad 
 {\Omega\over 2} \left\{ e^{-i\theta} \nabla^2 [e^{i\theta} (1 + \delta)] - 
   P(\delta) \right\} + e^{-i\theta} \eta/a ,
\nonumber 
\end{eqnarray}
where $P(\delta)= |r| \delta (1 + \delta) (2 + \delta) $.
Separating real and imaginary parts we obtain
\begin{eqnarray}
{\partial \theta\over \partial t} (1 + \delta) &= &
   {\Omega\over2}\left[ \nabla^2\theta (1 + \delta) + 
   2  \nabla \theta \cdot \nabla \delta\right] + \eta_\theta,
\\
{\partial \delta\over \partial t} &=& 
   - {\Omega\over2} \left[ (\nabla \theta)^2 (1 + \delta) 
   - \nabla^2 \delta + P(\delta) \right] + \eta_\delta,
\nonumber
\end{eqnarray}
where 
$\eta_\theta({\bm x},t)$ and $\eta_\delta({\bm x},t)$ 
have both probability distribution proportional to 
\begin{equation}
   \exp\left[ - {a^2\over 2 \Omega} 
    \int dt d{\bm x}\, \eta_\#({\bm x},t)^2\right].
\end{equation}
Now, we consider the evolution of $\theta$ and make the following
assumptions.
\begin{itemize}
\item[(i)] We set $1 + \delta \approx 1$. This is quite natural
as we expect $\delta$ to make small fluctuations around zero. 
\item[(ii)] We assume that $\delta$ and $\theta$ spatial 
fluctuations are uncorrelated, so
that  
\begin{equation}
\langle \nabla \theta \cdot \nabla \delta \rangle = 0.
\end{equation}
\end{itemize}
These two assumptions are equivalent to those made in the heuristic
argument.  Essentially, we are decoupling the fluctuations of the
modulus of the magnetization from its angular precession. If these two
conditions hold, we obtain the linear equation
\begin{eqnarray}
{\partial \theta\over \partial t} &= &
   {\Omega\over2}\nabla^2\theta + \eta_\theta.
\end{eqnarray}
We consider a box of size $L_1\times L_2\times L_3$, $V = L_1 L_2
L_3$, and define
\begin{eqnarray}
&&\theta({\bm k},t) = \int_V dx e^{i{\bm k}\cdot {\bm x}} \theta({\bm x},t), \\
&&\theta({\bm x},t) = {1\over V} \sum_{\bm k} e^{-i{\bm k}\cdot {\bm x}} 
     \theta({\bm k},t) .
\end{eqnarray}
We obtain
\begin{eqnarray}
{\partial \theta({\bm k},t)\over \partial t} &= &
   -{\Omega k^2 \over2}\theta({\bm k},t) + \eta_\theta({\bm k},t),
\end{eqnarray}
where the probability distribution of $\eta_\theta({\bm k},t)$ is 
proportional to
\begin{equation}
\exp\left[ - {a^2\over 2 V\Omega} 
    \int dt \sum_{\bm k} |\eta_\theta|^2\right].
\end{equation}
The solution is 
\begin{equation}
\theta({\bm k},t) = e^{-\Omega k^2 t/2} 
   \int_0^t e^{\Omega k^2 s/2} \eta_\theta({\bm k},s) ds,
\end{equation}
where we assume $\theta({\bm k},0) = 0$. If we average over the 
random noise, we obtain
\begin{equation}
\langle \theta({\bm k},t)^* \theta({\bm k},s) \rangle = 
   {V\over a^2 k^2} \left(
   e^{-\Omega k^2 |t-s|/2} - e^{-\Omega k^2 (t+s)/2} \right).
\end{equation}
If we now define the average angle 
\begin{equation}
\Theta(t) = {1\over V} \int d{\bm x}\, \theta({\bm x},t),
\end{equation}
it follows
\begin{equation}
\langle \Theta(t) \rangle = 0 \qquad
\langle \Theta(t) \Theta(s) \rangle = {\Omega \over 2 V a^2} (t + s - |t-s|),
\end{equation}
or, equivalently,
\begin{equation}
\langle [\Theta(t) - \Theta(s)]^2 \rangle = {\Omega \over a^2 V} |t-s|.
\end{equation}
This shows that the average angle changes on time scales of order $V$. 
Let us now consider a ``wall" quantity. If 
${\bm x} = (x_1,x_2,x_3)$ we consider 
\begin{equation}
   \delta \theta_\perp(x_3,t) = 
  {1\over L_1 L_2}\sum_{x_1,x_2} [\theta({\bm x},t)  - \Theta(t)]
\end{equation} 
and 
\begin{equation}
   G(t,s,r = x_3 - y_3) = 
   \langle \delta \theta_\perp(x_3,t) 
           \delta \theta_\perp(y_3,t)  \rangle.
\end{equation}
In the large-time limit, $t,s\to \infty$ at fixed $\tau = |t-s|$, i.e., at 
equilibrium, we can write it as 
\begin{eqnarray}
G(t,s,r) &=& {1\over a^2 V} \sum_{k_3\not=0} 
  {e^{-\Omega k^2_3 \tau/2}\over k^2_3} e^{i k_3 r} \nonumber \\
&=& {L_3 \over 4 \pi^2 a^2 L_1 L_2} 
    \sum_{n\not=0} {e^{-n^2 \widetilde{\tau}} \over n^2 }
     e^{2 \pi i n \widetilde{r}}, 
\end{eqnarray}
where 
\begin{equation}
   \widetilde{\tau} = 2 \pi^2 \Omega \tau/L_3^2 \qquad\qquad
   \widetilde{r} = r/L_3.
\end{equation} 
There are, therefore, two different regimes. If $L_1 = L_2 = L_3 = L$
(cubic case) fluctuations are small (their variance decreases as
$1/L$) and have a typical time scale of order $L^2$. If instead, $L_1
= L_2 = L$ and $L_3 \propto L^2$, fluctuations are finite, but occur
on significantly longer time and space scales $\tau \sim L^4$ and $r
\sim L^2$. Note that, for $\tau$ large, the correlation function
decreases exponentially, as $e^{-\tilde{\tau}}$.

At this point we can compute the autocorrelation function of the
magnetization. Assuming that the size fluctuations are not relevant,
we can write
\begin{equation}
\langle {\bm M}(s)\cdot {\bm M}(t) \rangle =  
   {a^2\over V^2} 
  \int d{\bm x} d{\bm y} \langle 
   e^{i[\theta({\bm x},s) - \theta({\bm y},t)]}\rangle.
\end{equation}
In a cubic geometry, $\theta({\bm x},t) = \Theta(t) + O(1/\sqrt{L})$ so that 
\begin{eqnarray}
&& \langle {\bm M}(s)\cdot {\bm M}(t) \rangle \approx
   a^2 \langle e^{i[\Theta(s) - \Theta(t)]} \rangle = 
\\
&& \qquad a^2 e^{-{1\over2} \langle[\Theta(s) - \Theta(t)]^2\rangle}  =
   a^2 \exp\left( - {\Omega \over 2 a^2 V} |t-s|\right).
\nonumber 
\end{eqnarray}
The autocorrelation function has therefore a typical exponential shape
with an autocorrelation time that scales as $V = L^3$. In the 
asymmetric case we should take into account the fluctuations along 
the longitudinal direction. Assuming a decoupling of the modes we obtain 
\begin{eqnarray}
&& \langle {\bm M}(s)\cdot {\bm M}(t) \rangle \approx
a^2 \exp\left( - {\Omega \over 2 a^2 V} |t-s|\right) \times 
\nonumber \\
&& \qquad {1\over L_3}
  \int_{-L_3/2}^{L_3/2} dr
    \exp[-G(t,s,r)/2].
\end{eqnarray}
In this case the behavior is more complex: in particular, we expect the
amplitude of the time decay to be smaller than in the cubic
case. However, the typical time dependence is still of order $V$, i.e.,
of order $L^4$.

\section{Numerical results}
\label{numerical}

We now present numerical results obtained by MC simulations of the
Heisenberg model, to check the off-equilibrium scaling ansatzes put
forward in Sec.~\ref{offequisec}. We implement the protocol described
in Sec.~\ref{protocol}.  We start from equilibrium configurations at
an initial value $h_i<0$ at $t_i<0$.  Then, the system evolves at
fixed temperature by means of a heat-bath updating
scheme~\cite{heatbath}.  The time unit is a sweep of the whole
lattice, that is a heat-bath update at all sites.  The magnetic field
is changed according to Eq.~(\ref{ht}) every sweep, incrementing $t$
by one.  The off-equilibrium relaxational dynamics ends at $t=t_f>0$,
corresponding to a finite $h_f=1/32$.  This procedure is repeated
several times, averaging the observables at fixed time $t$.  We will
first consider the behavior at $T_c$, then at $T < T_c$.

\subsection{Off-equilibrium scaling at $T_c$}
\label{nofftc}

\begin{figure}[tbp]
\includegraphics*[scale=\graphicscale]{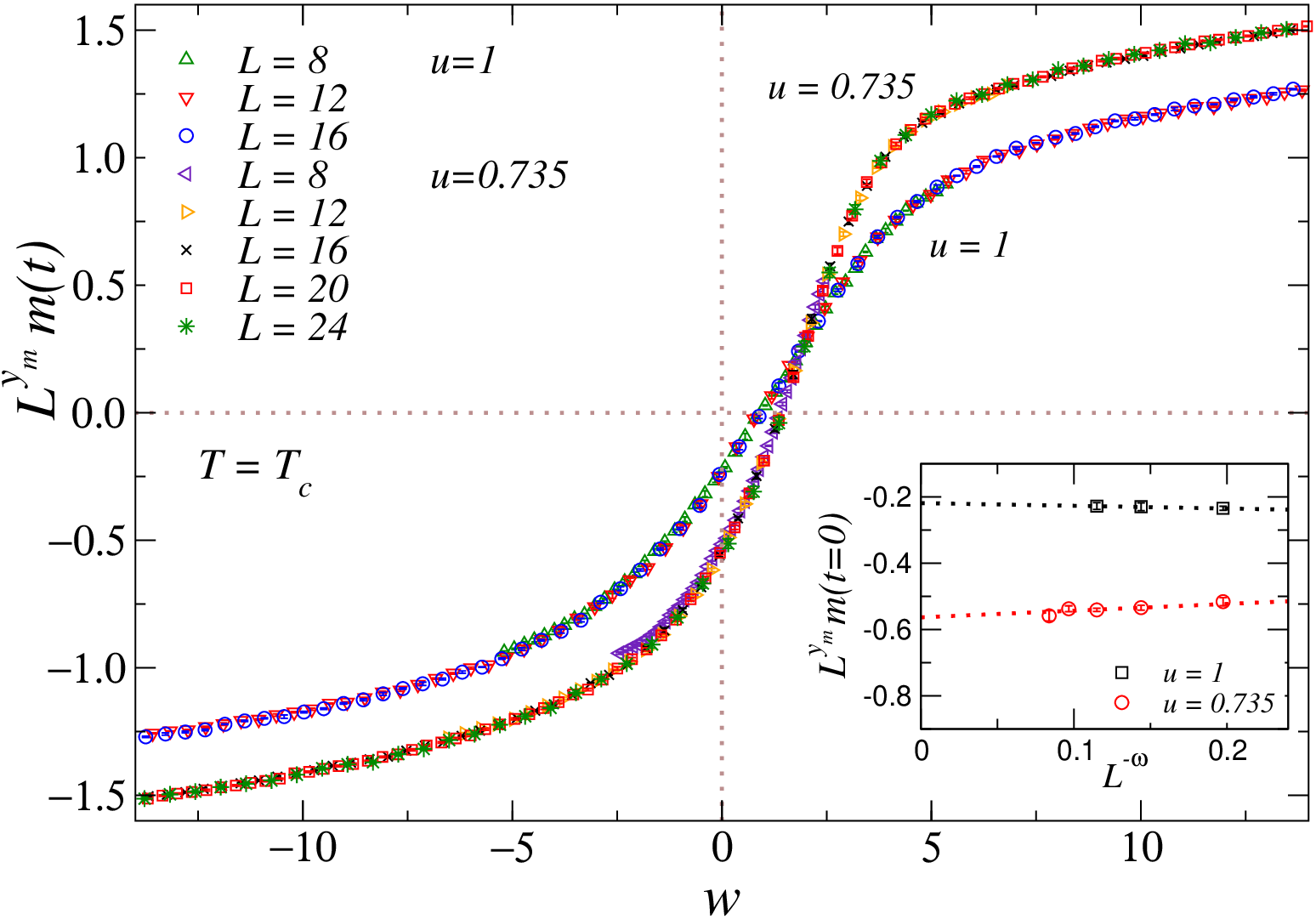}
\caption{(Color online) The rescaled magnetization $L^{y_m}
  m(t,t_s,L)$ during the off-equilibrium dynamics for cubic lattices
  at the critical point $T_c\approx 1.443$.  We report data for
  $u=0.735$ and $u=1$, versus $w$, as defined in Eqs.~(\ref{udef}) and
  (\ref{wdef}). We use $\kappa=0.2222$ and $\kappa_t = 0.449$, see
  Eq.~(\ref{heiofexp}).  The inset shows $L^{y_m} m(t=0)$ vs
  $L^{-\omega}$ with $\omega=0.78$, which is the expected behavior of
  the scaling corrections at $T_c$. }
\label{scaltcr}
\end{figure}

We first verify the off-equilibrium scaling relation (\ref{mcont}) for the
magnetization at the critical point $T_c$. The relevant exponents for
the Heisenberg $N=3$ model are
\begin{equation}
y_m =0.5189(2),\quad \kappa = 0.2222(5),\quad  \kappa_t = 0.449(1),
\label{heiofexp}
\end{equation}
obtained using the estimates $\eta=0.0378(3)$ and $z=2.02(1)$.
In Fig.~\ref{scaltcr} we show the product $L^{y_m} m$ at
fixed $u=t_s^\kappa/L$ versus $w=t/t_s^{\kappa_t}$, for 
$u\approx 0.735$ and $u=1$. The results clearly approach
$u$-dependent scaling curves with increasing $L$, nicely supporting
the off-equilibrium scaling relation (\ref{mcont}).  
Scaling corrections are expected to be
controlled by the leading irrelevant operator, thus they should decay
as $L^{-\omega}$ with $\omega\approx 0.78$. The data shown in the
inset of Fig.~\ref{scaltcr} are consistent with this prediction.

\begin{figure}[tbp]
\includegraphics*[scale=\graphicscale]{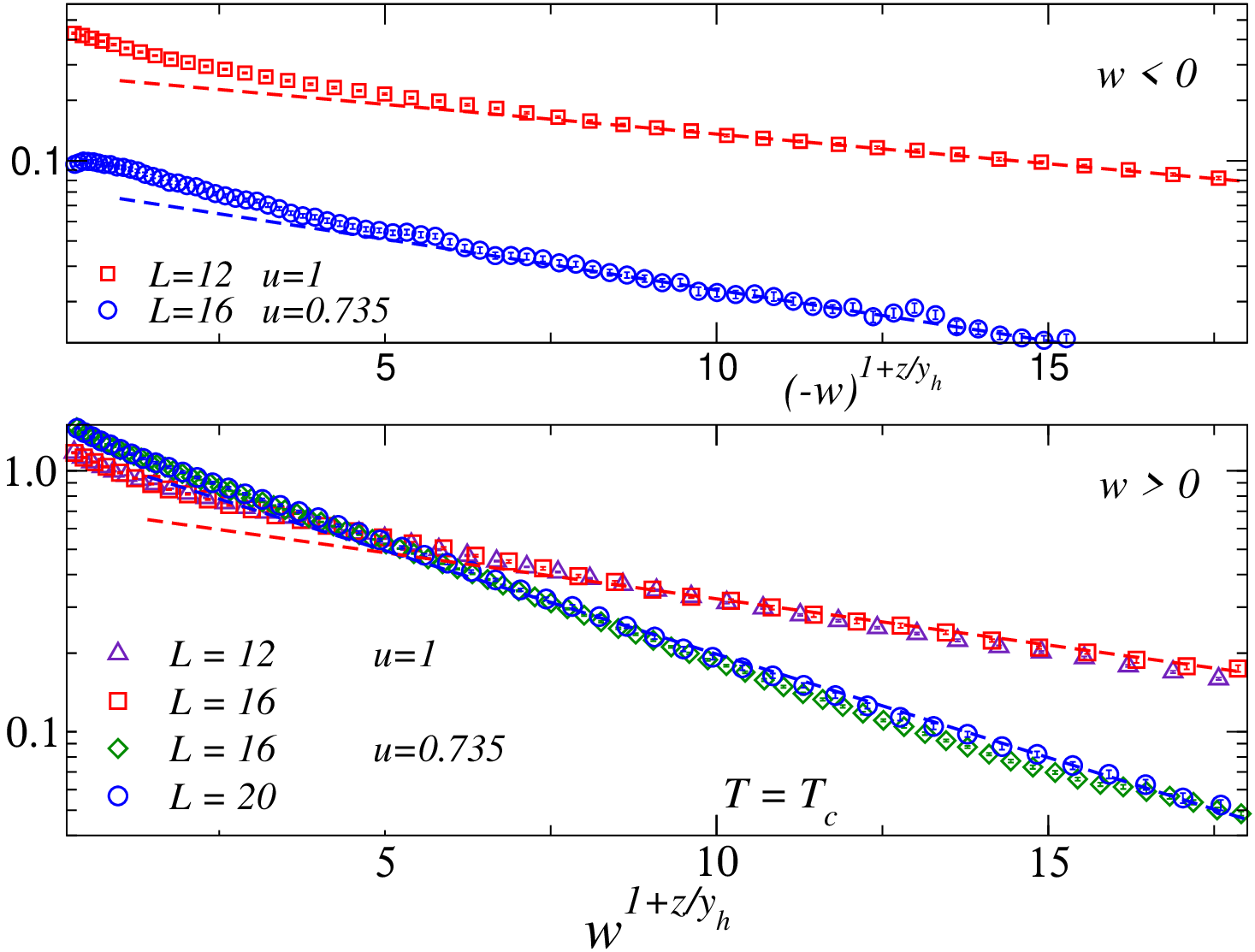}
\caption{(Color online) Semilogarithmic plot of $Q_\pm$, as defined in
  Eq.~(\ref{newqu}), for two values of $u$. We consider $w <
  0$ (top) and $w > 0$ (bottom). Data are consistent with the 
  exponential decay predicted by Eqs.~(\ref{asyexp}) and (\ref{asyexp2}).
 The straight lines are only meant to guide the eye.
}
\label{tcrexpappr}
\end{figure}

The numerical results are also consistent with the asymptotic
behaviors reported in Eqs.~(\ref{asyexp}) and (\ref{asyexp2}).  To
verify this explicitly we first determine the constant $a$, that
appears in Eq.~(\ref{F-largew}), by considering the large-$|w|$
behavior. For both values of $u$ we have investigated, and both for
$w\to+\infty$ and $w\to-\infty$, we obtain the same value $a \approx
0.745$, with a relative error that should be less than 1\%.  Then, we
consider the quantity
\begin{eqnarray}
Q_\pm(u,w) = 1 - {L^{y_m} m (t,t_s,L)\over f_{\pm \infty}(w u^{-y_h})}
\label{newqu}
\end{eqnarray}
where the subscript $\pm$ refers to positive and negative values of
$w$.  It is plotted in Fig.~\ref{tcrexpappr}, showing
that the data are consistent with the exponential decay predicted by
Eqs.~(\ref{asyexp}) and (\ref{asyexp2}).

\subsection{Off-equilibrium scaling for $T<T_c$}
\label{nofftmtc}

\begin{figure}[tbp]
\includegraphics*[scale=\graphicscale]{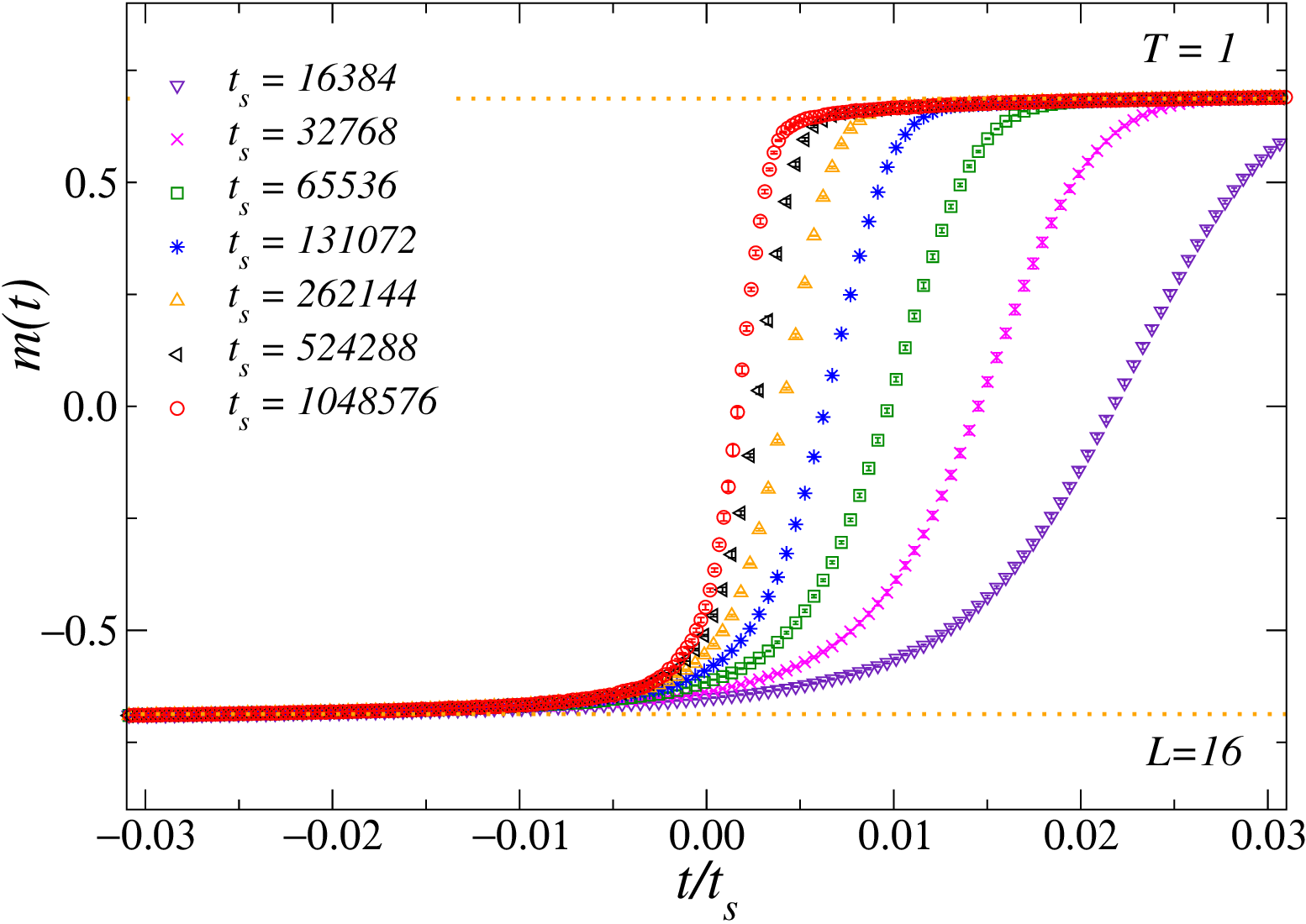}
\caption{(Color online) The magnetization as a function of $t/t_s$ for
  cubic systems of size $L=16$, for $T=1$ and several values of $t_s$.
  The dotted lines correspond to $m = \pm m_0(T=1)$ with
  $m_0(T=1)\approx 0.668$, where $m_0(T)$ is the equilibrium spontaneous 
  magnetization defined in Eq.~(\ref{m0t}).  }
\label{plain}
\end{figure}

We now provide numerical evidence for the off-equilibrium scaling
relation (\ref{fm}) along the first-order transition line $T<T_c$.
The results for the equilibrium static and dynamic FSS of
Sec.~\ref{eqbeh} lead to the following values for the exponents
$\kappa$ and $\kappa_t$:
\begin{eqnarray}
&&\kappa = 1/6,\quad \kappa_t = 1/2 \qquad ({\rm cubic}),
\label{cubicsy}\\
&&\kappa = 1/8,\quad \kappa_t = 1/2 \qquad ({\rm cylinder}),
\label{cylindersy}
\end{eqnarray}
where we used $y_h=3,\;z_m=3$ and $y_h=4,\;z_m=4$ for cubic and
anisotropic cylinder-like systems, respectively.  

One may qualitatively understand why $z_m$ is the relevant dynamical
exponent to be used to compute $\kappa$ and $\kappa_t$ in the
low-temperature case. For $h < 0$ the total magnetization is aligned
with $\bm H$. As $h$ changes sign, $\bm M$ should change its
direction. As the system is magnetized, this is achieved by means of a
rotation of $\bm M$, which takes a time of order $V$ to be
accomplished. During this time interval the system is out of
equilibrium and hence, the relevant time scale of the off-equilibrium
dynamics scales as $V = L^{z_m}$, as we have used in the discussion.

To verify the theoretical predictions, we have carried out simulations
at $T = 1$. Fig.~\ref{plain} shows estimates of the magnetization as a
function of $t/t_s$ for cubic systems of linear size $L=16$ and
several values of $t_s$.  For negative values of $t$, when
$t/t_s\lesssim -0.01$ say, $m(t)$ is approximately equal to $-m_0(T)$,
the spontaneous magnetization determined in equilibrium simulations at
$h =0$.  The system is approximately in equilibrium, indicating that
these values of $t_s$ are sufficiently large compared with the
relaxation time.  As the transition point $h=0$ is approached, the
system is not able to adapt itself to the changes of the magnetic
field.  We note that the magnetization at $t=0$ is always negative and
it increases with increasing $t_s$.  With increasing $t_s$, the
behavior around $t=0$ becomes sharper and sharper with respect to
$h(t)=t/t_s$, to reconstruct the equilibrium discontinuity at $h=0$.
For $t_s\to \infty$ at fixed $L$ we expect $m(t=0)\to 0$, since the
magnetization for $h=0$ vanishes at equilibrium due to the O(3)
symmetry.  We now show that this behavior around $t=0$ can be
described by the off-equilibrium scaling relation (\ref{fm}) for
$t\approx 0$.

\begin{figure}[tbp]
\includegraphics*[scale=\graphicscale]{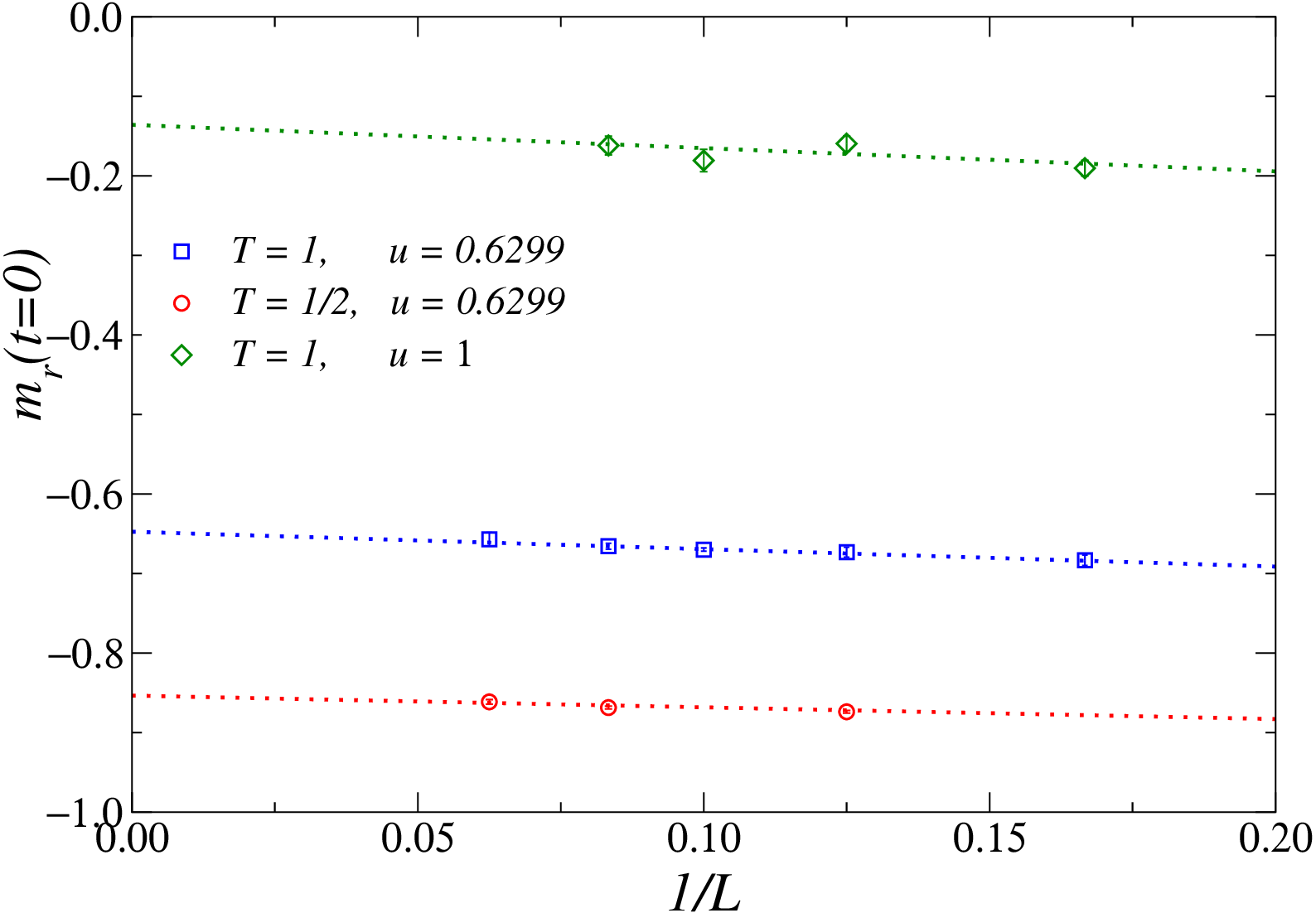}
\caption{(Color online) The renormalized magnetization $m_r$ at $t=0$ 
  for cubic $L^3$ systems as a function of $1/L$. We report data for
  some values of $T<T_c$ and $u\equiv t_s^{\kappa}/L$ with $\kappa=1/6$.
}
\label{scal0}
\end{figure}

\begin{figure}[tbp]
\includegraphics*[scale=\graphicscale]{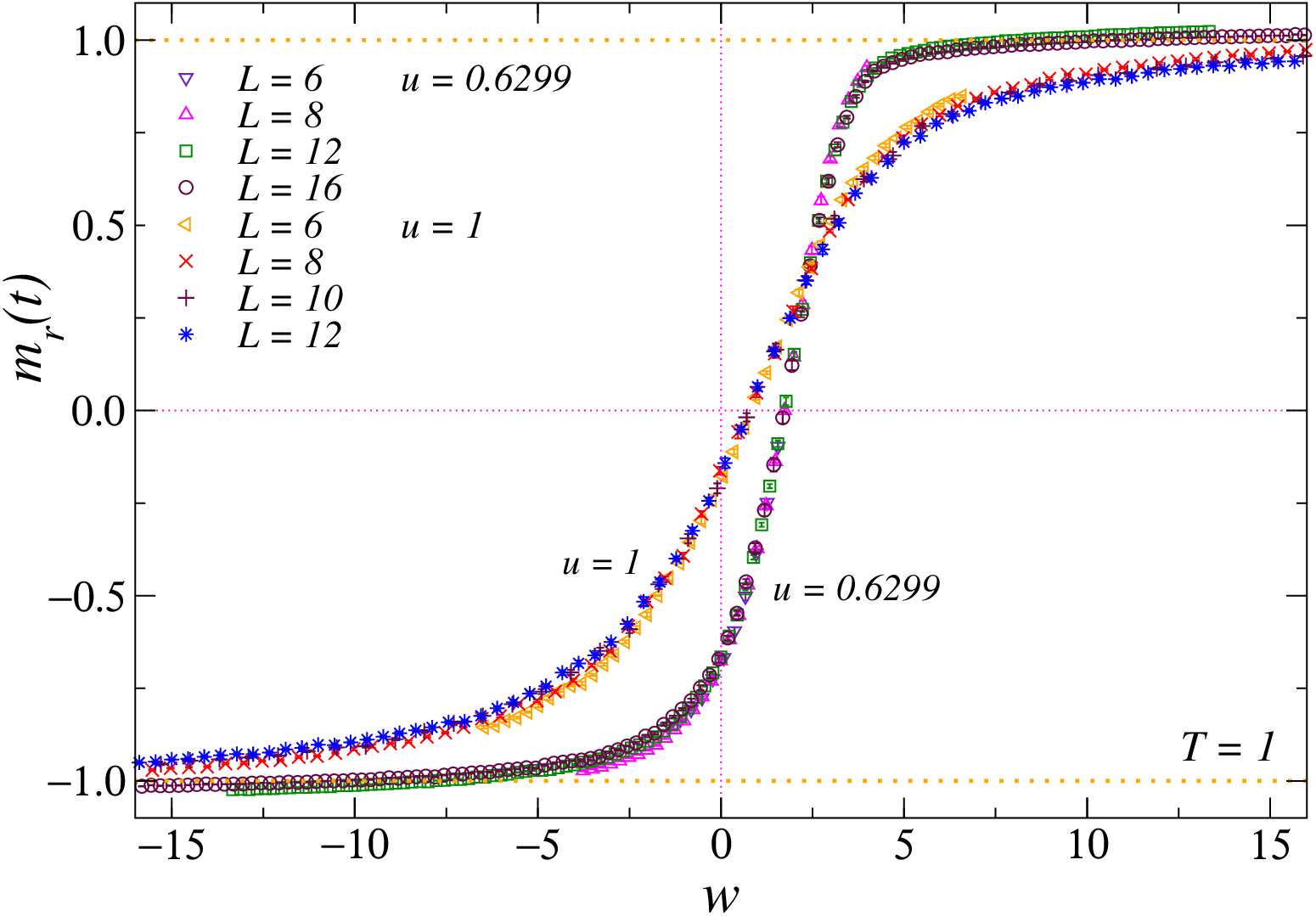}
\caption{(Color online) Scaling of the renormalized magnetization for
  cubic systems, as a function of $w\equiv t/t_s^{\kappa_t}$ with
  $\kappa_t=1/2$.  We report data for $T=1$ and two values of the
  scaling variable $u=t_s^\kappa/L$ with $\kappa=1/6$.  }
\label{scalT1}
\end{figure}

\begin{figure}[tbp]
\includegraphics*[scale=\graphicscale]{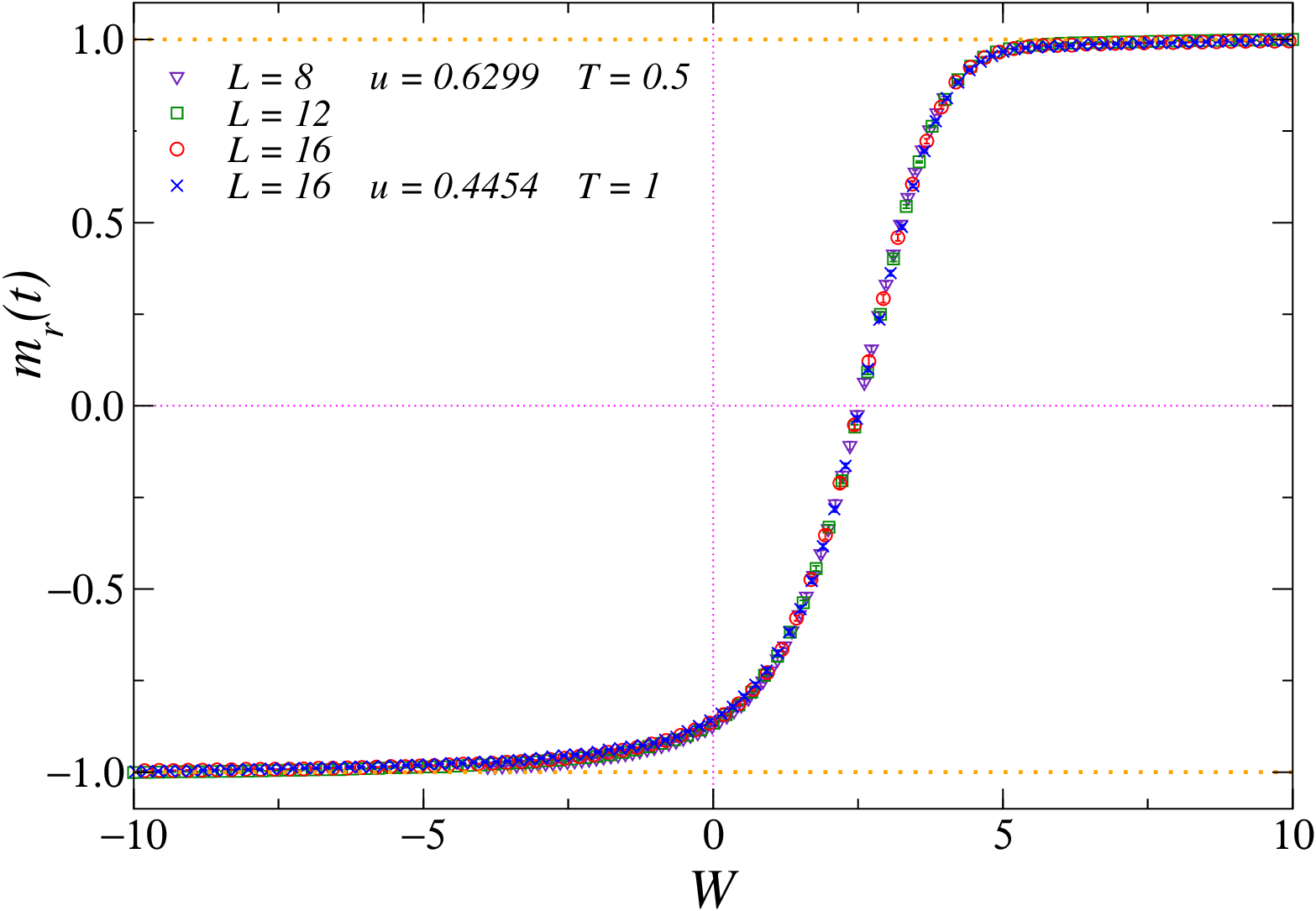}
\caption{(Color online) Scaling of the renormalized magnetization at $T=0.5$
  and $T = 1$ at the same value of $U = 0.6299$ (correspondingly 
  $u = 0.4454$ for $T=1$ and $u = 0.6299$ for $T=0.5$) 
  as a function of $W$ ($w = W$ for $T =0.5$ and $w = 0.9 W$ for $T=1$).
}
\label{scalT1o2}
\end{figure}

In order to verify that $u=t_s^\kappa/L$ with $\kappa=1/6$ is the
correct scaling variable for cubic systems, we note that
Eq.~(\ref{fm}) for $t=0$ implies the relation
\begin{equation}
m_r(0,t_s,L) \approx g(u).
\label{mrer0}
\end{equation}
We expect $g(u)$ to be an increasing function of $u$, and in
particular that $g(u\to 0)=-1$, and $g(u\to\infty) = 0$ due to the
fact that at equilibrium $\langle {\bm M} \rangle = 0$ for ${\bm
  H}=0$.  Eq.~(\ref{mrer0}) implies that data at $t=0$ and fixed
$u=t_s^\kappa/L$ must converge to nontrivial $u$-dependent values with
increasing $L$.  Fig.~\ref{scal0} shows data at some fixed values of
$u$.  They appear to converge to nontrivial values, supporting the
predicted asymptotic behavior, with corrections which decay as $L^{-1}$,
as expected.  Note that the different values obtained at $T=1$ and
$T=1/2$ for the same value of $u$ do not contradict universality,
because universality implies the same scaling function $g$ apart from a
normalization of the argument, see below.  

The scaling with respect to $w=t/t_s^{\kappa_t}$ at fixed values of
$u$ is supported by the results reported in Figs.~\ref{scalT1} and
\ref{scalT1o2}, where we show results for $T=1$ and $T=0.5$,
respectively.  In all cases the data at fixed values of $u$ approach
an asymptotic function of the scaling variable $w$, as predicted by
the off-equilibrium scaling theory.  

We expect that the scaling behavior (\ref{fm}) is universal with
respect to changes of the temperature $T$ as long as $T<T_c$. The
scaling curves should be the same apart from trivial normalizations of
the scaling variables $u$ and $w$.  To make universality more evident,
we define new variables $U = c_u(T) u$ and $W = c_w(T) w$ so that
\begin{equation}
   m_r(t,t_s,L,T) = \hat{F}_m(U,W),
\end{equation}
where $\hat{F}_m(U,W)$ is universal and $T$ independent. All
temperature dependence is encoded in the two nonuniversal functions
$c_u(T)$ and $c_w(T)$. They are uniquely specified only if two
normalization conditions are given. For example, one can fix the
scaling function for two particular values of $U$ and $W$, or specify 
$c_u(T)$ and $c_w(T)$ at a given temperature.  In the
following we require $c_u(T) = c_w(T) = 1$ for $T=0.5$. To perform a
universality check using data at $T=0.5$ and $T=1$, we should
determine $c_u(T)$ and $c_w(T)$ for $T=1$. To determine the former
quantity, we consider data at $t = 0$ (correspondingly $w = 0$) and
note that the data for $T=1$ at $u = 0.4454$ and $T=0.5$ at $u =
0.6299$ both give $m_r(t=0) \approx -0.86$ for $L\to \infty$.  This
implies $c_u(T=1)\approx 0.6299/0.4454 = 1.414$. To determine
$c_w(T=1)$ we consider the results for the same value of $U$ (we take $U =
0.6299$) and require the data to collapse once plotted as a function
of $W$. We obtain $c_w(T=1) \approx 1.1$. The results for $T=1$
and $T=0.5$ at the same value of $U$ are shown in Fig.~\ref{scalT1o2}
as a function of $W$: all data fall onto the same scaling curve,
nicely confirming universality.

\begin{figure}[tbp]
\includegraphics*[scale=\graphicscale]{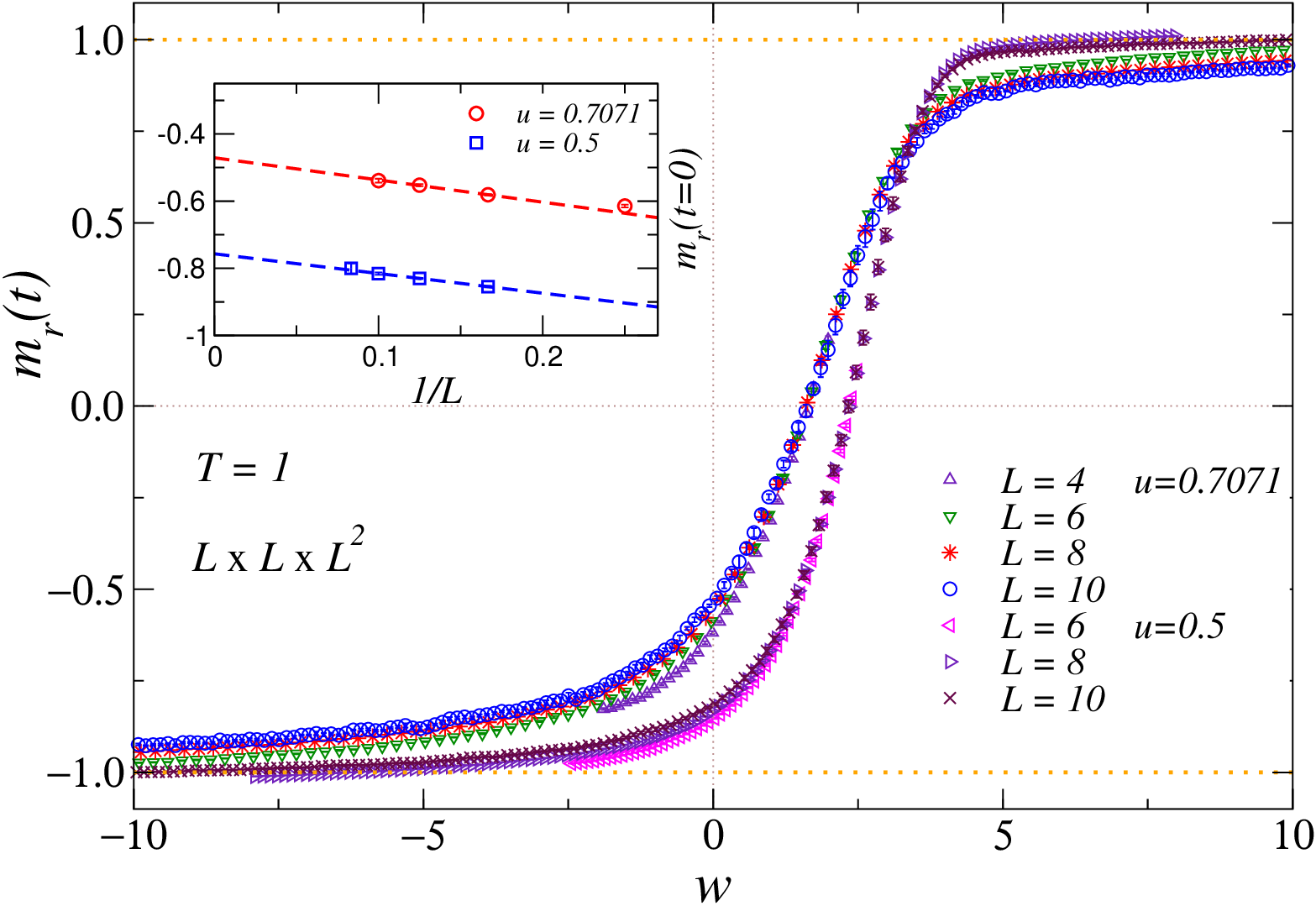}
\caption{(Color online) Estimates of the renormalized magnetization
  for anisotropic lattices of size $L^2\times L_\parallel$ with
  $L_\parallel = L^2$.  We report data for $T=1$ and
  $u=t_s^\kappa/L=0.7071,\,0.5$ (with $\kappa=1/8$), versus the
  scaling variable $w\equiv t/t_s^{\kappa_t}$ (with
  $\kappa_t=1/2$). The inset shows $m_r(t=0)$ vs $1/L$, which is the
  expected behavior of the scaling corrections. }
\label{scalan}
\end{figure}

Analogous results are obtained in the case of cylinder-like systems,
when using the corresponding scaling exponents given in
Eq.~(\ref{cylindersy}).  In Fig.~\ref{scalan} we show the renormalized
magnetization for a system of size $L\times L\times L^2$.  Results at
fixed $u=t_s^\kappa/L$ with $\kappa=1/8$ are plotted versus
$w=t/t_s^{\kappa_t}$ with $\kappa_t=1/2$. The data appear to collapse
toward scaling curves, confirming the correctness of the scaling
Ansatz (\ref{fm}) with the exponents (\ref{cylindersy}).

\section{Hysteresis phenomena}
\label{hysteresissec}

\begin{figure}[tbp]
\includegraphics*[scale=\graphicscale]{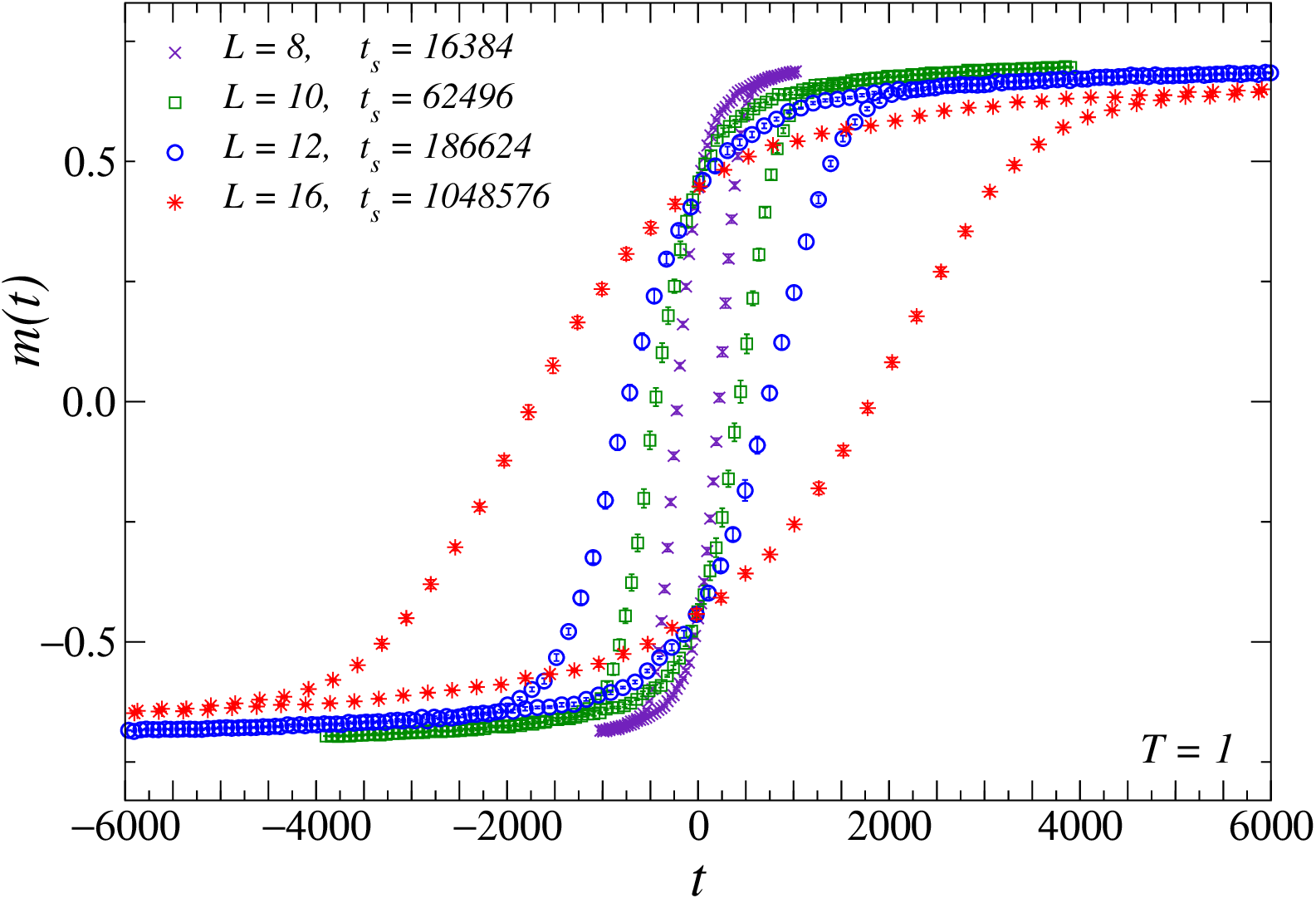}
\caption{(Color online) Hysteresis loops of the magnetization for
  cubic systems at $T=1$ versus the {\em time} parameter $t$, for
  several values of $L$ and $t_s$.  We consider a round-trip protocol:
  first $t$ increases from $t_i < 0$ to $t_f>0$ (correspondingly, we
  have $h_i = -1/16$ and $h_f = 1/16$), then it decreases back to $t_i
  < 0$.  }
\label{plainhyst}
\end{figure}

\begin{figure}[tbp]
\includegraphics*[scale=\graphicscale]{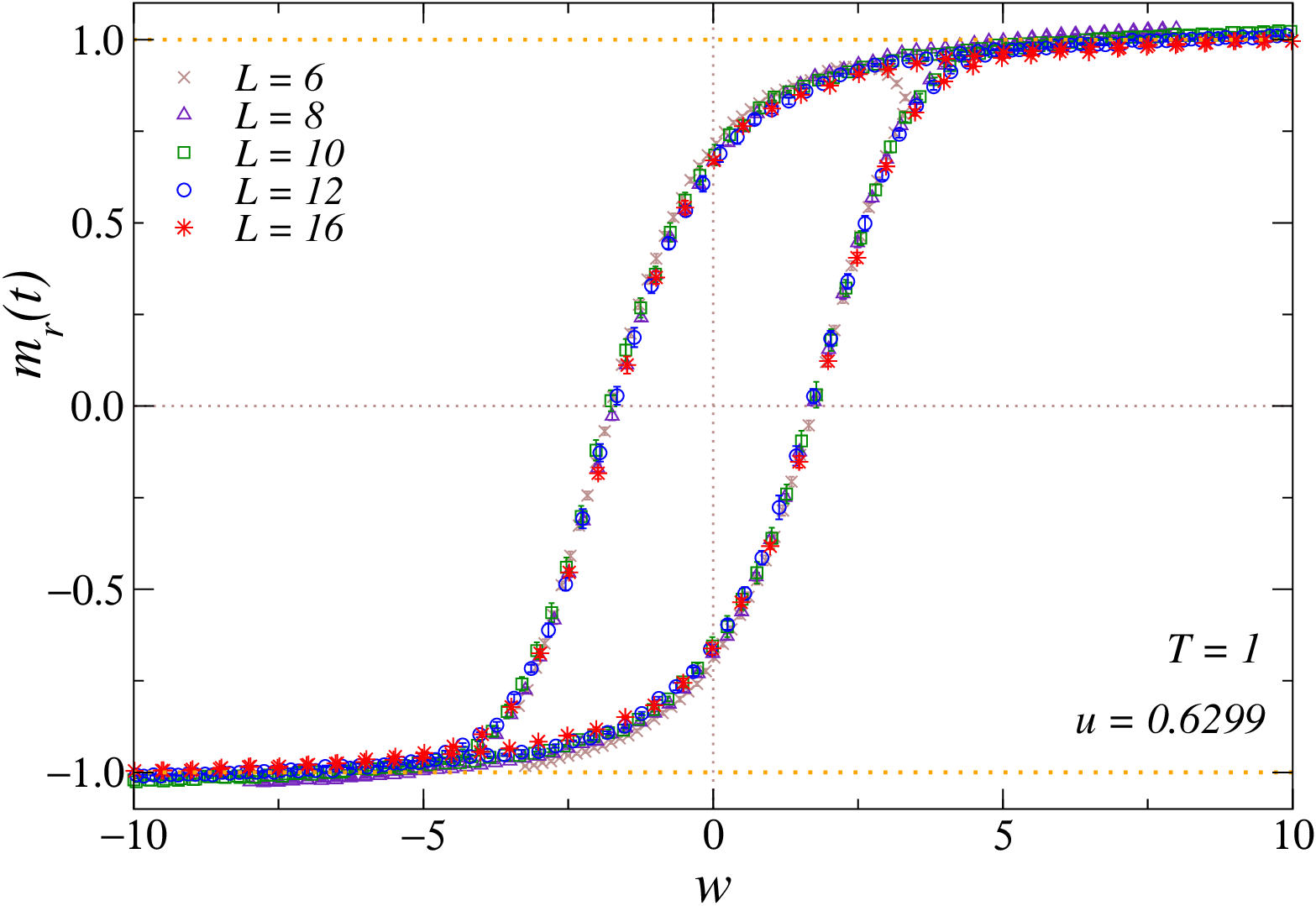}
\caption{(Color online) Hysteresis loop of the renormalized
  magnetization $m_r$ defined in Eq.~(\ref{renmag}) for cubic $L^3$
  systems at $T=1$. We report data at fixed $u\approx 0.6299$ versus
  $w$.  We consider a round-trip protocol: first $t$ increases from
  $t_i < 0$ to $t_f>0$ (correspondingly, we have $h_i = -1/16$ and
  $h_f = 1/16$), then it decreases back to $t_i < 0$.  }
\label{hysteresis}
\end{figure}

\begin{figure}[tbp]
\includegraphics*[scale=\graphicscale]{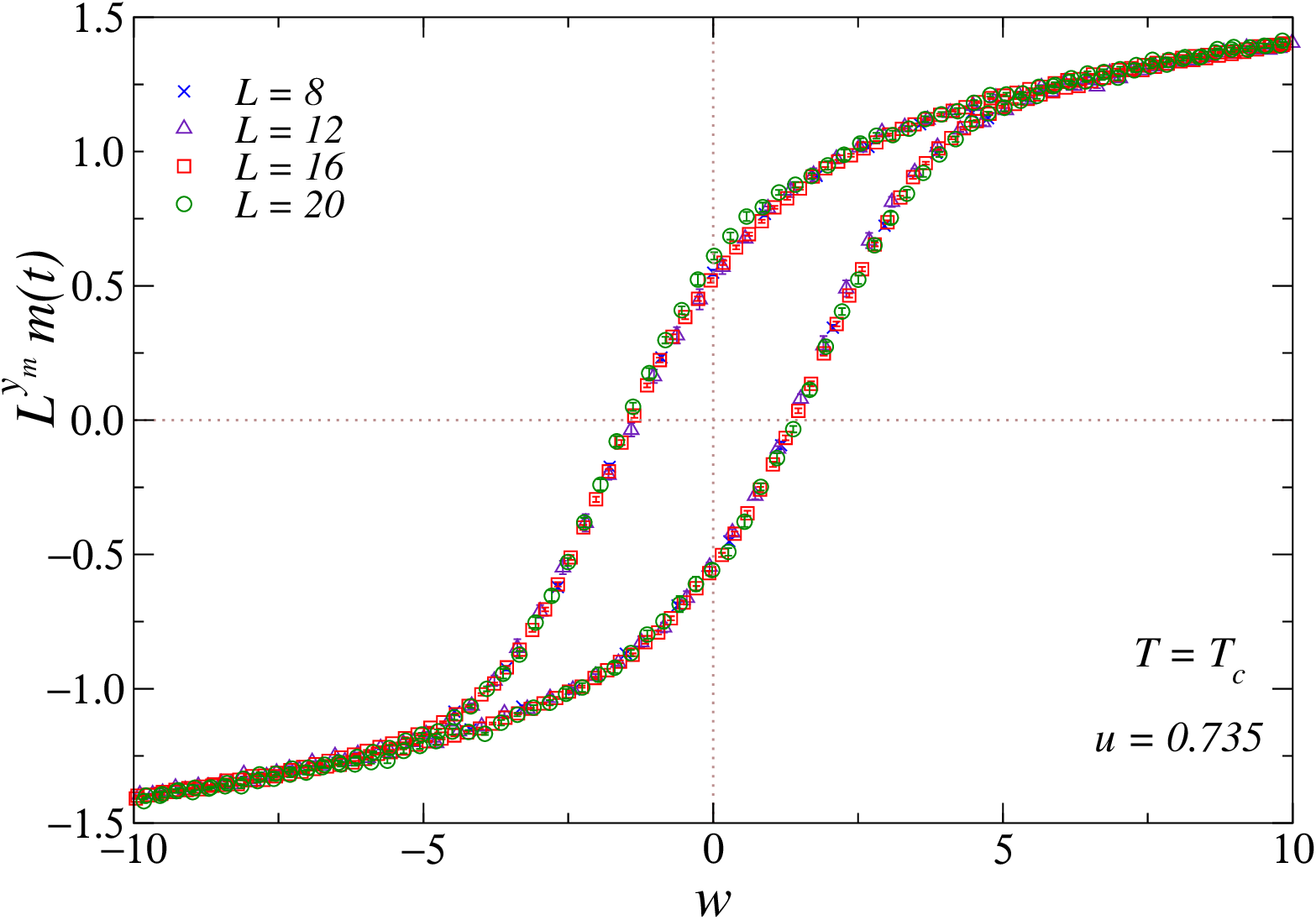}
\caption{(Color online) Plot of $L^{y_m}m(t)$ for cubic $L^3$ systems
  at the critical point $T_c$.  We report data at fixed $u\approx
  0.735$ versus $w$.  We consider a round-trip protocol: first $t$
  increases from $t_i < 0$ to $t_f>0$ (correspondingly, we have $h_i =
  -1/16$ and $h_f = 1/16$), then it decreases back to $t_i < 0$.  }
\label{hysteresisc}
\end{figure}

\begin{figure}[tbp]
\includegraphics*[scale=\graphicscale]{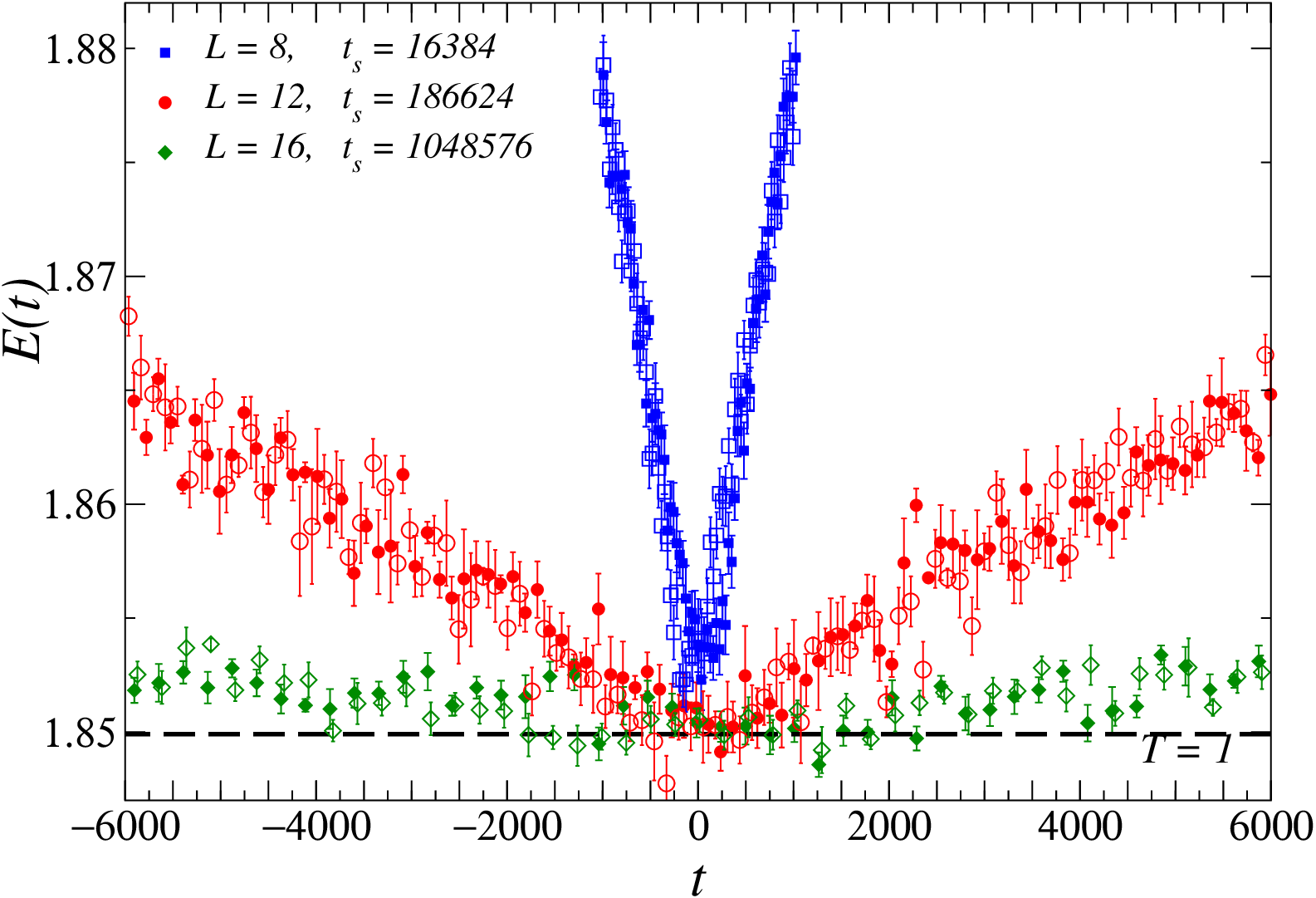}
\caption{(Color online) Time dependence of the energy density at $T=1$
  along the same round-trip protocols considered in
  Fig.~\ref{plainhyst}: first $t$ increases from $t_i < 0$ to $t_f>0$
  (full symbols), correspondingly, we have $h_i = -1/16$ and $h_f =
  1/16$, then it decreases back to $t_i < 0$ (open symbols).  There is
  no evidence of hysteresis within the precision of our data. 
  The dashed line corresponds to the
  equilibrium value for $h=0$.  }
\label{plaine}
\end{figure}

As shown in the previous sections, the system is unable to
reach equilibrium when it goes through the transition point at $h=0$.
To quantify the departure from equilibrium we can consider protocols
in which the magnetic field is slowly increased from $h_i < 0$ to $h_f
> 0$ and then it is decreased back again to $h_i < 0$.  In this case
the magnetization shows a hysteresis loop, whose area 
\begin{equation}
A_h = -\oint dt\, m(t)
\label{aharea}
\end{equation}
provides a quantitative indication of how far the system is out of
equilibrium. 

In Fig.~\ref{plainhyst} we show some examples of hysteresis loops for
the magnetization for cubic systems at $T=1$. Here we start at $t_i <
0$ ($h_i = -1/16$), increase $t$ until $h= h_f = 1/16$, then decrease
$t$ back to $t_i$.  The arguments presented in the previous sections
imply that also the hysteresis loops have a scaling behavior.  Scaling
plots are shown in Fig.~\ref{hysteresis} for $T=1$ and in
Fig.~\ref{hysteresisc} at the critical point. It should be noted that,
while the magnetization shows a clear hysteresis cycle, there is no
evidence of such a phenomenon for the energy. In Fig.~\ref{plaine} we
show the time dependence of the energy density at $T=1$ obtained using
the same round-trip protocols considered in Fig.~\ref{plainhyst}.
Within the precision of our data, there is no evidence of hysteresis.

In practice, since time is discretized in our MC simulations we
measure $m(t)$ at discrete values $t_j$ of $t$, the area enclosed by
the hysteresis loop of the magnetization can be computed using the
area estimator
\begin{eqnarray}
B_h \equiv \Delta 
\sum_j \left[
m(t_j,t_s,L)_{h_f\to h_i}  - m(t_j,t_s,L)_{h_i\to h_f} \right],\;\;
\label{area}
\end{eqnarray}
where $\Delta \equiv t_{j+1}-t_j$ is the time interval between two
measurements.

Using the scaling relations (\ref{mcont}) and (\ref{fm}) for the
magnetization, we can express the area in terms of the scaling
functions $F_m(u,v)$ and $F_m^{({\rm inv})}(u,w)$ of the reverse
process in which time decreases.  At the critical point $T_c$, we
obtain 
\begin{eqnarray}
A_h &\approx& L^{-y_m} \,t_s^{\kappa_t} \int_{-\infty}^{\infty}
dw\; \left[ F_m^{({\rm inv})}(u,w)  - F_m(u,w) \right] \quad\label{ahcont}\\
&=& - L^{y_m} \,t_s^{\kappa_t} \int_{-\infty}^{\infty}
dw\; \left[ F_m(u,w)  + F_m(u,-w) \right],
\nonumber 
\end{eqnarray}
where relation (\ref{fminv}) has been used to arrive at the second line.
Using the asymptotic relation (\ref{asyexp}), it is immediate to show
that the integral (\ref{ahcont}) is finite.  We can therefore define a
universal scaling function ${\cal A}(u)$ associated with the area of
the hysteresis loop:
\begin{eqnarray}
&&A_h \approx L^{-y_m} \, t_s^{\kappa_t}  {\cal A}(u),\label{ahcont2}\\ 
&&{\cal A}(u) = - \int_{-\infty}^{\infty}
dw\; \left[ F_m(u,w)  + F_m(u,-w) \right].
\nonumber 
\end{eqnarray}
We expect ${\cal A}(u)$ to be a decreasing function of $u$, and in
particular ${\cal A}(u)\to 0$ in the static limit $u\to\infty$.  The
scaling relation (\ref{ahcont2}) can also be written as
\begin{equation}
A_h\approx L^{z_m-y_m} u^{z_m} {\cal A}(u),
\label{rewri}
\end{equation}
where $z_m - y_m = 3/2 + (c-1/2)\eta$, see Eqs.~(\ref{ymyhct}) and
(\ref{zcthree}). For the Heisenberg model we have $z_m-y_m \approx
1.50$, so that $A_h$ increases with $L$ at fixed $u$.

Proceeding analogously, in the low-temperature phase $T<T_c$ we obtain
\begin{eqnarray}
&&A_h \approx m_0(T) \; t_s^{\kappa_t} {\cal A}(u), 
\label{ahlowt}\\ 
&&{\cal A}(u) = - \int_{-\infty}^{\infty}
dw\; \left[ F_m(u,w)  + F_m(u,-w) \right],
\nonumber
\end{eqnarray}
where $F_m$ is the scaling function entering Eq.~(\ref{fm}).
Eq.~(\ref{ahlowt}) can be also written as $A_h\approx L^{z_m} u^{z_m}
{\cal A}(u)$.  We stress that the scaling behavior of the hysteresis
loop area does not depend on the chosen values $h_i<0$ and $h_f>0$, as
already discussed in Sec.~\ref{offequisec}.

\begin{figure}[tbp]
\includegraphics*[scale=\graphicscale]{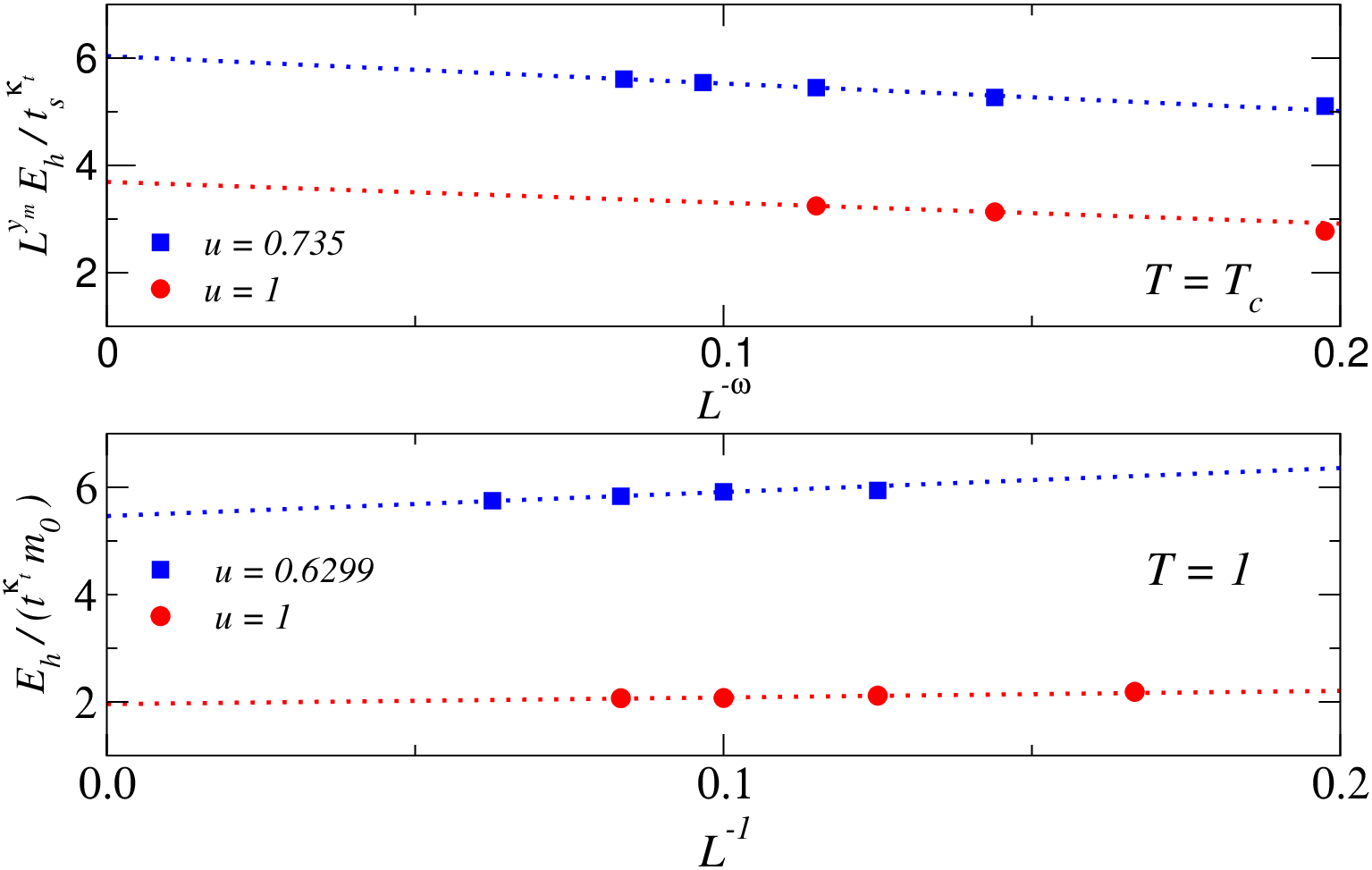}
\caption{(Color online) Hysteresis loop area at $T=T_c$ (top) and at
  $T=1$ (bottom).  We report $t_s^{-\kappa_t} L^{y_m} E_h$ versus
  $L^{-\omega}$ in the case of $T=T_c$ (here we use $\kappa=0.2222$
  and $\kappa_t=0.449$), and $t_s^{-\kappa_t} m_0(T)^{-1} E_h$ versus
  $L^{-1}$ in the $T=1$ case (here we use $\kappa=1/6$ and
  $\kappa_t=1/2$).  The dotted lines correspond to linear fits of the
  data for the largest available sizes.  }
\label{scalhyst}
\end{figure}

By exploiting the parity symmetry leading to the relation
(\ref{fminv}), the hysteresis loop area can be estimated by only using
data of the one-way protocol from $h_i<0$ to $h_f>0$, introduced in
Sec.~\ref{protocol}.  This allows us to define an improved estimator
$E_h$
\begin{eqnarray}
E_h(t_s,L;T) &\equiv& \Delta \sum_j [m(-t_j,t_s,L) - m(t_j,t_s,L)]  
\nonumber\\
&=&-2\Delta \sum_j m(t_j,t_s,L),  
\label{impest}
\end{eqnarray}
which has a smaller statistical errors than $B_h$ defined in 
Eq.~(\ref{area}).  In Fig.~\ref{scalhyst} we report $E_h$ for $T=1$
and $T=T_c$ at some fixed values of $u$. In order to check the
scaling behaviors (\ref{ahcont}) and (\ref{ahlowt}), we plot
$t_s^{-\kappa_t} L^{y_m} E_h$ versus $L^{-\omega}$ in the case of
$T=T_c$, and $t_s^{-\kappa_t} m_0(T)^{-1} E_h$ versus $L^{-1}$ in the
$T=1$ case.  The results linearly extrapolate to constant values,
supporting the predicted off-equilibrium scaling behaviors
(\ref{ahcont}) and (\ref{ahlowt}).  The estimator $B_h$ 
gives consistent results.

\section{Conclusions}
\label{conclu}

We consider systems with a continuous global symmetry in the presence
of slowly-varying time-dependent external fields. More specifically,
we focus on the 3D O($N$) vector model with $N\ge 2$ coupled to a
time-dependent spatially uniform magnetic field ${\bm H}(t)$.  We
assume ${\bm H}(t) = h(t)\,{\bm e}$, where $h(t)=t/t_s$, $t_s$ is a
time scale, and $\bm e$ is a constant unit $N$-component vector.  In
practice, the dynamics starts from equilibrium configurations at an
initial value $h_i<0$ at time $t_i<0$. Then, the magnetic field is
slowly varied, up to a time $t_f>0$ corresponding to a finite $h_f>0$.
For $t\approx 0$, i.e. when the magnetic field $h$ changes sign, the
system is no longer in equilibrium, for any temperature $T \le T_c$,
where $T_c$ is the critical temperature.

In the adiabatic limit $t_s \to \infty$, the off-equilibrium behavior
turns out to be universal. By using scaling arguments, we derive a
general scaling theory for the dynamics across the transition line $h
= 0$. In particular, the magnetization shows universal scaling
behaviors in terms of appropriate combinations of the time $t$, the
time scale $t_s$, and the finite size $L$, in the large-$L,\,t_s,\,t$
limit. Heuristic scaling arguments allow us to identify the relevant
scaling variables, which can be expressed as $t_s^\kappa/L$ and
$t/t_s^{\kappa_t}$, where $\kappa$ and $\kappa_t$ are appropriate
exponents.  In other words, as the system slowly moves through the
transition line $h = 0$, the dynamics is controlled by a new length
scale $\xi\sim t_s^\kappa$ and by a time scale $\tau\sim
t_s^{\kappa_t}$.  We stress that this is not specific of the
continuous transition at $T_c$. The same scaling theory, but with
different exponents, applies in the low-temperature phase $T<T_c$,
where the system undergoes a first-order transition at $h = 0$.
 
The exponents $\kappa$ and $\kappa_t$ depend on the
equilibrium static and dynamic exponents which describe the
equilibrium static and dynamic FSS behavior of the system at fixed
temperature.  Static FSS is recovered as a particular limit
of the off-equilibrium scaling ansatzes. For $T < T_c$ the exponents
do not vary with the temperature, but depend on the boundary
conditions and on the geometry of system. In particular, the exponents
for cubic $L\times L \times L$ systems are different from those
appropriate for anisotropic cylinder-like systems of size $L\times L
\times L_\parallel$ with $L_\parallel\gg L$.  Such a phenomenon does
not occur at $T_c$: here the exponents do not depend on the shape and
on the boundary conditions (however, scaling functions do depend on
these features).

We present MC numerical results for the 3D Heisenberg ($N=3$) model.
We use a heat-bath updating scheme, which is a particular example of a
purely relaxational dynamics (sometimes named as model-A dynamics
\cite{HH-77}).  The results provide a robust numerical evidence of the
off-equilibrium scaling theory we have put forward, both at the
critical point $T_c$ and for $T<T_c$. Note that the general theory
should hold for any $N\ge 2$ in three dimensions, and it should also
apply to other types of dynamics, provided one uses the appropriate
value for the dynamic exponent.

We also discuss the hysteresis phenomena which generally arise in
off-equilibrium conditions when round-trip protocols are used.  We
consider again the time-dependent external field $h(t)=t/t_s$, but now
we vary $t$ first from $t_i<0$ to $t_f>0$, and then back from $t_f$ to
$t_i$.  The area enclosed by the hysteresis curve of the magnetization
may be considered as a measure of how far the system is from
equilibrium.  The hysteresis loop area satisfies a nontrivial scaling
behavior at and below $T_c$, as a function of the scaling variable
$t_s^\kappa/L$.

In the case of discrete symmetries, such as the Ising ($N=1$) model or
the Potts models, the off-equilibrium behavior in the low-temperature
phase may be even more complex than the one discussed
here.  As already suggested by studies of classical and quantum
systems with discrete symmetry~\cite{PF-83,CNPV-14,CPV-15,PV-15}, the
off-equilibrium behavior is expected to be particularly sensitive to
the boundary conditions. In particular, drastically different
behaviors may be observed, depending on the presence/absence of an
interface in the system.

As mentioned in the introduction, off-equilibrium phenomena arising
from slow changes of model parameters through phase transitions are of
great theoretical and experimental interest. They have been
investigated in several experiments on different physical systems,
see, e.g.,
Refs.~\cite{CDTY-91,BCSS-94,BBFGP-96,Ruutu-etal-96,CPK-00,CGMB-01,MMR-02,
  MPK-03,CGM-06,MMARK-06,SHLVS-06,WNSBDA-08,GPK-10,CWCD-11,
  Chae-etal-12,MBMG-13,EH-13,Ulm-etal-13,Pyka-etal-13,LDSDF-13,
  Corman-etal-14,NGSH-15,Braun-etal-15,DWGGP-16}.  Most investigations
have been performed at continuous classical and quantum
transitions. We have shown here that analogous scaling behaviors 
can be observed at first-order transitions, with some peculiar features.  We
believe that theoretical and experimental investigations of these
issues may lead to a substantial progress in the understanding of 
dynamic phenomena at first-order classical and quantum transitions,
which are observed in many different physical systems, from magnetic
to cold-atom systems, both at finite temperature and in the
zero-temperature limit.

\end{document}